\documentclass[12pt,letter,doublespace]{article}
\usepackage{amssymb}
\usepackage{natbib,graphicx,setspace,lscape,longtable}
\usepackage{natbib,epsfig,graphicx}
\usepackage{amsmath,amsthm,amssymb}
\usepackage{algorithmic}
\usepackage{color,xcolor}
\usepackage{threeparttable}
\usepackage{changes}

\usepackage[figuresright]{rotating}
\usepackage{rotating}

\usepackage{booktabs}

\usepackage{hyperref}

\usepackage[linesnumbered,boxed,ruled,commentsnumbered]{algorithm2e}

\RequirePackage[mathlines, displaymath]{lineno}
\usepackage{multirow}
\usepackage{graphicx}
\usepackage{subfig}
\usepackage{threeparttable}
\DeclareGraphicsExtensions{.eps, .jpg, .pdf}
\bibpunct{(}{)}{;}{a}{,}{,}

\newcommand{\scsection}[1]
    {\begin{center}
        {\bf\large #1}
    \end{center}
    \vspace{-0.15 cm}
}

\def\beqr{\begin{eqnarray}}
\def\eeqr{\end{eqnarray}}
\def\beqrs{\begin{eqnarray*}}
\def\eeqrs{\end{eqnarray*}}
\def\bep{\begin{prop}}
\def\eep{\end{prop}}
\def\bc{\begin{center}}
\def\ec{\end{center}}

\newcommand{\bs}{\mathbf{s}}

\newcommand{\Z}{\mathbf{Z}}

\newcommand{\bH}{\mathbf{H}}

\newcommand{\bP}{\mathbf{P}}

\newcommand{\trans}{^{\mbox{\tiny{T}}}}

\numberwithin{equation}{section}

\def \bec{\begin{center}}
\def \enc {\end{center}}
\def \bee {\begin{eqnarray*}}
\def \ene {\end{eqnarray*}}

\def \bear{\begin{array}}
\def \enar{\end{array}}

\def \bs{\begin{slide}}
\def \es{\end{slide}}

\newcommand{\zzeta}{\boldsymbol{\zeta}}

\newcommand{\ggamma}{\boldsymbol{\gamma}}
\newcommand{\ttheta}{\boldsymbol{\theta}}

\newcommand{\SSigma}{\boldsymbol{\Sigma}}

\newcommand{\xxi}{\boldsymbol{\xi}}

\newcommand{\PPsi}{\boldsymbol{\Psi}}
\newcommand{\eeta}{\boldsymbol{\eta}}
\newcommand{\llambda}{\boldsymbol{\lambda}}
\newcommand{\LLambda}{\boldsymbol{\Lambda}}
\newcommand{\UUpsilon}{\boldsymbol{\Upsilon}}

\newcommand{\bbb}{{\bf b}}

\newcommand{\bB}{{\bf B}}

\newcommand{\bT}{{\bf T}}
\newcommand{\bV}{{\bf V}}
\newcommand{\bU}{{\bf U}}

\newcommand{\bX}{{\bf X}}

\newcommand{\bZ}{{\bf Z}}
\newcommand{\bW}{{\bf W}}
\newcommand{\bM}{{\bf M}}

\newcommand{\bI}{{\bf I}}
\newcommand{\bC}{{\bf C}}

\numberwithin{equation}{section}  % number the equations by section

\newtheorem{thm}{Theorem}[section]
\newtheorem{lem}{Lemma}[section]
\newtheorem{rem}{Remark}[section]
\newtheorem{cor}{Corollary}[section]

\newtheorem{prop}{Proposition}[section]
\renewcommand{\baselinestretch}{1.5}
\begin{document}
%\linenumbers

\title{Composite Estimation for Quantile Regression Kink  Models with Longitudinal Data}
\author{{Chuang Wan\footnote{ Email: {wanchuang@stu.xmu.edu.cn}.}}\\
Xiamen University, China}
%\date{}
\maketitle{}

\begin{abstract}
Kink model is developed to analyze the data where the regression function is two-stage linear but intersects at an unknown threshold. In quantile regression with longitudinal data, previous work assumed that the unknown threshold  parameters or kink points are heterogeneous across different quantiles. However, the location where kink effect happens tend to be the same across different quantiles, especially in a region of neighboring quantile levels. Ignoring such homogeneity information may lead to efficiency loss for estimation. In view of this, we propose a composite estimator for the common kink point by absorbing information from multiple quantiles. In addition, we also develop a sup-likelihood-ratio test to check the kink effect at a given quantile level. A test-inversion confidence interval for the common kink point is also developed  based on the quantile rank score test. The simulation study shows that the proposed composite kink estimator is more competitive with the least square estimator and the single quantile estimator. We illustrate the practical value of this work through the analysis of a body mass index and blood pressure data set.
\end{abstract}


\pagestyle{plain}
\setcounter{page}{1}

\section{Introduction}%\label{sec:1}
Quantile regression,  as a useful complement to mean regression, provides a systematic tool to describe the conditional  distribution of a response given covariates and is more robust to outliers and heavy-tailed errors. Due to these merits, quantile regression has been  extensively applied  in diverse fields and also popularized in kinds of data types. One of the  important data type in statistic and biostatistics  is the longitudinal data, where the measurements on the same subject are repeatedly observed. So the observations within one subject are generally correlated and ignoring such correlation structure may bring  statistical analysis biases. In the past two decades, a great deal of literatures have been performed to study the quantile regression for longitudinal data, see  for example \cite{koenker2004quantile}, \cite{tang2011empirical}, \cite{leng2014smoothing}, \cite{tang2015improving} and \cite{wang2019copula}.

The literatures mentioned above always assume that the regression coefficients are constant on the whole domain of predictors. However, such stability of coefficients may be violated in some applications. For example, \cite{li2015quantile} studied the cognitive decline for patients with Alzheimer disease (AD) and found that cognitive function declined as normal aging  in the early preclinical stage of AD and then accelerated with the progress of disease. To capture this distinctive feature,  a quantile regression kink model for longitudinal data is developed in their paper.  Kink regression, as a special threshold model, describes a situation where the threshold effect happens at an unknown change point in one covariate while the regression function is  continuous all over the domain of predictors. Such regression has been widely applied in cross-sectional data \citep{li2011bent},  time series data \citep{hansen2017regression} and  binary data \citep{fong2017model}, partly due to its balance between  interpretability of linear model and the flexibility of nonparametric regression.

In kink models, the threshold parameter or the kink point denotes the location where the slope of a threshold predictor changes is usually of great research interest.  \cite{li2015quantile} proposed a profiled estimation strategy to estimate model parameters by assuming that kink points are heterogeneous across different quantiles. Thus the kink points are actually estimated at each given quantile level separately. However, in some cases, the kink parameters at different quantiles, especially in neighboring quantiles tend to be the same. For example, in our empirical analysis, body mass index shows different  kink effects on blood pressure at different quantiles, but the kink points appears to occur around the same location at a certain region.
The estimators obtained  at a single quantile  may not be efficient.  Although \cite{zhang2017composite} studied the composite change point estimation  in independent and identically distributed data, proper estimation and inference procedures for composite estimator still have not been established  for longitudinal data.

In this paper, we consider a joint regression analysis of multiple quantiles for  kink  regression in longitudinal study. Compared to the literature, we make the following four main contributions. First, we propose a two-stage profile estimation strategy to estimate the common kink point by combining the information from different quantiles. We demonstrate that the composite estimator is more efficient than a single quantile  analysis through simulation study. Second, to further check the kink effect at a given quantile, we construct a sup-likelihood-ratio test and a wild blockwise bootstrap procedures is developed to characterize the limiting distribution. Third, as the traditional Wald-type confidence interval for the kink estimator does not perform well, a test-inversion set based on the quantile rank score test in longitudinal data is developed to improve the limiting performance. Fourth, we apply the proposed composite method to the  longitudinal body mass index and blood pressure data and get some interesting findings. Our method can provide a more informative  analysis tool for biostatistics.

The rest of this paper is organized as follows. In Sect 2, we describe the detailed estimation procedures for the composite quantile kink regression with longitudinal data, and derive the asymptotic properties. In Sect 3, we make statistical inference on the kink estimators including the kink effect test and constructing the confidence interval. A series of simulation studies is conducted in Sect 4 to evaluate the finite sample performance of proposed methods and an application of blood pressure data analysis is illustrated in Sect 5. Sect 6 concludes this paper. The technical proofs are given in the Appendix. The R code implementing all methods is available at author's github: \url{https://github.com/ChuangWAN1994/CQRCPM}.

%their method can not be applied to longitudinal data directly as they did not consider the correlation  within one subjects.

%In kink models, the threshold parameter or the kink point denotes the location where the slope of a threshold predictor changes is usually of great research interest.  Li (2015) proposed a profiled estimation strategy to estimate all paramters by assuming that kink points are heterogeneous across different quantiles.

% Fong (2018) developed a fast bootstrap  algorithm to substantially improve the coverage probability of parameters when the model is mis-specified. Hidalgo (2019) proposed a quasi-likelihood-ratio test to check whether there experiences a jump or kink at the kink point.

%One issue with using quantile methods is that the resulting estimator may not be efﬁcient if estimation is obtained at a single quantile level

\section{Model and  Asymptotic Property}
\subsection{Model setup and estimation}\label{sec:21}
Suppose that we have $N$ individuals or subjects and for $i$th individual, it is measured $n_i$ times. So there are totally $n=\sum_{i=1}^nn_i$ observations. We denote $Y_{ij}$ as the $i$th response for $j$th individual, $X_{ij}$ as a bounded scalar covariate with thresholding effect and $\Z_{ij}$ as a $q$-dimensional additional covariates of interest. For any given quantile index $\tau\in(0,1)$, define the $\tau$th quantile of $Y_{ij}$ given $\bW_{ij}$  as $Q_Y(\tau|\bW_{ij})=F^{-1}(\tau|\bW_{ij})=\inf\{y:F(y|\bW_{ij})\}$ where $\bW_{ij}=(X_{ij},\Z_{ij}\trans)\trans$ and $F(\cdot|\bW)$ is the conditional cumulative density function of $Y$ given $\bW$.

We assume that the regressor $X$ has a continuous threshold effect on the response variable $Y$ at $K$ quantile levels $0<\tau_1<\cdots<\tau_K<1$, where $K$ is a finite integer. In this paper, we are interested in the following composite quantile regression for kink model  with longitudinal data:
\begin{eqnarray}\label{eq:cqr}
\begin{array}{c}
Q_Y(\tau;\eeta_{\tau_k},t|\bW_{ij})=\alpha_{\tau_k}+\beta_{1,\tau_k}(X_{ij}-t)I(X_{ij}\leq t)+\beta_{2,\tau_k}(X_{ij}-t)I(X_{ij}>t)+\bZ_{ij}\trans\ggamma_{\tau_k}\\
\text{for $i=1,\cdots,N$; $j=1,\cdots,n_i$ and $k=1,\cdots,K$}
\end{array}
\end{eqnarray}
where $\eeta_{\tau_k}=(\alpha_{\tau_k},\beta_{1,\tau_k},\beta_{2,\tau_k},\ggamma_{\tau_k}\trans)\trans$ are the regression coefficients at  $\tau_k$, $t$ is a common change point shared by  $K$ baseline models with different quantile levels and $I(A)$ is an indicator function, taking 1 when $A$ is true, otherwise 0. Obviously, the slope of $X_{ij}$ equals to $\beta_{1,\tau_k}$ when $X_{ij}$ is less than $t$, but turns into $\beta_{2,\tau_k}$ for values of $X_{ij}$ greater than $t$. Meanwhile, the slopes of $\Z_{ij}$ stay constant on the whole domain ares. Remark that the slope of $X_{ij}$ experiences a kink at $X_{ij}=t$ while the regression function $Q_Y(\tau;\eeta_{\tau_k},t|\bW_{ij})$ is everywhere continuous. Such phenomenon is generally referred to as kink effect or bent line effect.  The unknown parameter $t$ is therefore called change point, kink point or other terminologies.
The index set $\{\tau_k;k=1,\cdots,K\}$ are  user-specified. When $K=1$, Model (\ref{eq:cqr}) is degenerated to the standard  longitudinal kink model with a kink point, which has been studied by \cite{li2015quantile}. Here  we  focus on the composite estimator for the kink point $t$, which implies that the change point $t$ stays constant across $\tau_k$s.

Denote $\ttheta=(\eeta\trans,t)\trans$ and $\eeta=(\eeta_{\tau_1}\trans,\cdots,\eeta_{\tau_k}\trans)\trans$. The objective function for  estimating $\ttheta$ is
\begin{equation}\label{eq:obj}
S_n(\ttheta)=n^{-1}\sum_{k=1}^K\sum_{i=1}^N\sum_{j=1}^{n_i}\rho_{\tau_k}\{Y_{ij}-Q_Y(\tau_k;\eeta_{\tau_k},t|\bW_{ij})\}
\end{equation}
where $\rho_{\tau_k}(v)=v\{\tau_k-I(v<0)\}$ is the check loss function at level $\tau_k$. The standard estimator for $\ttheta$ is therefore given by
$$
\widehat{\ttheta}_n=\underset{\eeta\in\mathcal{B},t\in[M_1+\epsilon,M_2-\epsilon]}{\arg\min}S_n(\ttheta)
$$
where $\mathcal{B}\subset\mathbb{R}^{K(q+3)}$ is a compact set for $\eeta$,  $M_1$ and $M_2$ denotes the upper and lower bounds for $t$ and $\epsilon$ is a small positive number to avoid the edge effect. However, the objective function (\ref{eq:obj}) is  non-differentiable and non-smooth with respect to $t$, making the traditional convex optimization technique not applicable here. Inspired by \cite{li2015quantile} and \cite{zhang2017composite}, we adopt a two-stage profile estimation strategy to  minimize (\ref{eq:obj}).  The detailed procedures go as follows:

\textbf{Step 1.} Note that $S_n(\ttheta)$ is linear in $\eeta$ for a given candidate $t$. So the estimator for $\eeta$ conditional on $t$ can be estimated by %$\widehat{\eeta}_n(t)$.
\begin{equation}\label{eq:step1}
\widehat{\eeta}_n(t)=\underset{\eeta\in\mathcal{B}}{\arg\min}S_n\{\eeta(t),t\}.
\end{equation}
The minimization problem in (\ref{eq:step1}) becomes a standard linear quantile regression, which can be readily implemented by some existing convex optimization packages.  However, just as pointed by \cite{zhang2017composite}, for multiple quantiles estimation, there may exist such situation that the estimates at upper  quantile levels are smaller than that at lower  quantile levels, i.e. the crossing of quantile curves. Toward this end, we estimate $\eeta(t)$ by imposing a non-crossing constraint proposed by \cite{bondell2010noncrossing}. One can refer to their paper for more details about the crossing issues.

\textbf{Step 2.} Then the change point estimator $\widehat{t}_n$ is given by
\begin{equation}\label{eq:step2}
\widehat{t}_n=\underset{t\in[M_1+\epsilon,M_2-\epsilon]\cap(X_{n(2)},X_{n(n-1)})}{\arg\min}S_n\{\widehat{\eeta}_n(t),t\}
\end{equation}
where $X_{n(2)}$ and $X_{n(n-1)}$ are the 2nd and $(n-1)$th order statistics of $X_{ij}$. In the specific implementation, we adopt the optimization function ``\emph{optimize}" in \textbf{R} software to solve  (\ref{eq:step2}). The ultimate estimators for $\ttheta$ is therefore $\widehat{\ttheta}_n=(\widehat{\eeta}_n(\widehat{t}_n)\trans,\widehat{t}_n)\trans$.

\subsection{Large sample properties }
We now derive the asymptotic properties of $\widehat{\ttheta}_n$. Before, we first need to introduce some notations. Define the true parameters as  $\ttheta_0=(\eeta_0^\top,t_0)^\top$ and $u_{ij}^{(\tau_k)}=Y_{ij}-Q_Y(\tau;\eeta_{\tau_k},t|\bW_{ij})$ as the error term with $\tau_k$th conditional quantile being zero. Furthermore, we define two matrixes:  %define ${\LLambda}_n=E\left\{\frac{\partial^2S_n(\ttheta_0)}{\partial\ttheta\partial
%\ttheta}\right\}$ and ${\bf H}_n=E\left\{\frac{\partial S_n(\ttheta_0)}{S\partial\ttheta}\frac{\partial S_n(\ttheta_0)}{S\partial\ttheta^\top}\right\}$ at $\ttheta=\ttheta_0$. Thus
\begin{eqnarray}
{\bf H}_n=n^{-1}\sum_{k=1}^K\sum_{l=1}^K\left\{\begin{array}{ll}
\left[\sum_{i,j}\tau_k(1-\tau_k)
{\bf h}_k({\bf W}_{ij};\ttheta_0){\bf h}_k({\bf W}_{ij};\ttheta_0)\trans\right. & \\
\left.+\sum_i\sum_{j\neq j^{'}}\left(\delta_{ijj^{'}}^{(\tau_k)}-\tau_k^2\right){\bf h}_k({\bf W}_{ij};\ttheta_0)
{\bf h}_k({\bf W}_{ij^{'}};\ttheta_0)\trans\right] & (k=l)\\
&\\
\left[\sum_{i,j}(\tau_k\wedge\tau_l-\tau_k\tau_l)
{\bf h}_k({\bf W}_{ij};\ttheta_0){\bf h}_l({\bf W}_{ij};\ttheta_0)\trans\right. & \\
\left.+\sum_i\sum_{j\neq j^{'}}\left(\xi_{ijj^{'}}^{(\tau_k)}-\tau_k^2\right){\bf h}_k({\bf W}_{ij};\ttheta_0)
{\bf h}_l({\bf W}_{ij^{'}};\ttheta_0)\trans\right] & (k\neq l)\\
\end{array}
\right.
\end{eqnarray}
where  ${\bf h}_k({\bf W}_{ij};\ttheta)=({\bf 0}_{(k-1)(p+3)}^\top,{\bX}_{ij}(t)^\top,{\bf 0}^\top_{(K-k)(q+3)},-\beta_{1,\tau_k}I(X_{ij}\leq t)-\beta_{2,\tau_k}I(X_{ij}>t))^\top$,  ${\bX}_{ij}(t)=(1,(X_{ij}-t)I(X_{ij}\leq t),(X_{ij}-t)I(X_{ij}>t),{\bf Z}_{ij}^\top)^\top$; $\delta_{ijj^{'}}^{(\tau)}=P\left(u_{ij}^{(\tau)}<0,u_{ij^{'}}^{(\tau)}<0
\right)$ and $\xi_{ijj^{'}}^{(\tau_k,\tau_l)}=P\left(u_{ij}^{(\tau_k)}<0,u_{ij}^{(\tau_l)}<0\right)$. Define
\begin{eqnarray}\nonumber
\LLambda_n&=&n^{-1}\sum_{k=1}^K\sum_{i=1}^N\sum_{j=1}^{n_i}\frac{\partial}
{\partial\ttheta}\psi_{\tau_k}\{Y_{ij}-Q_Y(\tau_k;\eeta_{\tau_k},t|{\bf W}_{ij})\}{\bf h}_k({\bf W}_{ij};\ttheta)\Big|_{\ttheta=\ttheta_0}\\
&=&n^{-1}\sum_{k=1}^K\sum_{i=1}^N\sum_{j=1}^{n_i}f_{ij}^{(\tau_k)}\{Q_Y(\tau_k;\eeta_{\tau_k,0},t_0|{\bf W}_{ij})\}{\bf h}_k({\bf W}_{ij};\ttheta_0){\bf h}_k({\bf W}_{ij};\ttheta_0)\trans
\end{eqnarray}
where $\psi_{\tau_k}(v)=\tau_k-I(v\leq0)$ and $f_{ij}^{(\tau_k)} \{Q_Y(\tau_k;\eeta_{\tau_k,0},t_0|{\bf W}_{ij})\}$ is the probability density function of $Y_{ij}$ given ${\bf W}_{ij}$.

We make the following necessary regularity conditions:
\begin{itemize}
\item[(A1)]  The conditional distribution function $F_{ij,k}\equiv F(\tau_k|\bW_{ij})$ has first order derivative denoted by $f_{ij}^{(\tau_k)}(\cdot)$, which is uniformly bounded away from infinity at the point $F^{-1}(\tau_k|\bW_{ij})$ for all $i$, $j$ and $k$. The density $f_{ij}^{(\tau_k)}(\cdot)$ is Lipschitz continuous.
%for each $i,j$  and $k$ has a bound conditional Lebesgue density $f_{ij}^{(\tau_k)}$ that is Lipschitz continuous at the point $F^{-1}(\tau_k|\bW_{ij})$.
\item[(A2)] Threshold variable $X_{ij}$ is dense in the interval $[M_1,M_2]$ and has  a continuous and bounded  density function.
\item[(A3)] $\max_{i,j}\|\Z_{ij}\|=O(n^{1/4})$ and $n^{-1}\sum_{i,j}\|\Z_{ij}\|^3=O(1)$ as $n\rightarrow\infty$.
\item[(A4)] $\bH_n\rightarrow\bH$ and $\LLambda_n\rightarrow\LLambda$ as $n\rightarrow\infty$ where $\bH$ and $\LLambda$ are two positive definite matrices.
\item[(A5)] There exists a $S(\ttheta)$ such that $ES_n(\ttheta)\rightarrow S(\ttheta)$  as $n\rightarrow\infty$, which achieves a unique global minimum at true parameters $\ttheta_0$.
\end{itemize}
Assumption (A1) is standard in quantile regression. Assumptions (A2) and (A3) impose some conditions for threshold variable  $X_{ij}$ and additional covariates $\Z_{ij}$, which can also be found in \cite{li2015quantile}. Assumption (A5) ensures that the estimation is identifiable.

The following convergence result holds.
\begin{thm}\label{thm1}
Suppose the Assumptions (A1)-(A4) hold and given $\beta_{1,\tau_k}\neq\beta_{2,\tau_k}$ in Model (\ref{eq:cqr}), as $n\rightarrow\infty$, $\widehat{\ttheta}_n$ is a consistent estimator for $\ttheta_0$ and
$$
\sqrt{n}(\widehat{\ttheta}_n-\ttheta_0)\stackrel{d}{\longrightarrow}N({\textbf{0}},\SSigma)
$$
where $\SSigma=\LLambda^{-1}{\bf H}\LLambda^{-1}$.
\end{thm}

Moreover, we separately estimate $\LLambda$ and $\bH$ by plugging in $\widehat{\LLambda}_n$ and $\widehat{\bH}_n$ in which
\begin{eqnarray}
\widehat{\LLambda}_n=\left(\begin{array}{cc}
\widehat{\LLambda}_{n11} & \widehat{\LLambda}_{n12}\\
\widehat{\LLambda}^\top_{n12} & \widehat{\LLambda}_{n22}\\
\end{array}\right)
\end{eqnarray}
where $\widehat{\LLambda}_{n11}=-diag(\widehat{\llambda}_{n11,1},\cdots,
\widehat{\llambda}_{n11,K})$ is a $K(q+3)\times K(q+3)$ block diagonal matrix. For any $k=1,\cdots,K$,  $\widehat{\llambda}_{n11,k}$ is a $(q+3)\times(q+3)$ symmetric matrix given by
$$
\widehat{\llambda}_{n11,k}=n^{-1}\sum_{i=1}^N\sum_{j=1}^{n_i}\hat{f}^{(\tau_k)}_{ij}\{
Q_Y(\tau_k;\widehat{\eeta}_{\tau_k,n},\widehat{t}_n|\bW_{ij})\}\bX_{ij}(\widehat{t}_n)\bX_{ij}(\widehat{t}_n)\trans
%\bh_k(\bW_{ij};\widehat{\ttheta}_n)\bh_k(\bW_{ij};\widehat{\ttheta}_n)^\top,
$$
where $\hat{f}^{(\tau)}_{ij}(\cdot)$ is a consistent estimator for $f^{(\tau)}_{ij}(\cdot)$. In practical implementation, we estimate $f_{ij}^{(\tau)}(\cdot)$  by using the difference  quotient method of \cite{hendricks1992hierarchical}
$$
\hat{f}_{ij}^{(\tau)}\{Q_Y(\tau;\widehat{\eeta}_{\tau,n},\widehat{t}_n|\bW_{ij})\}=\max\left\{0,\frac{2\Delta_n}{Q_Y(\tau+\Delta_n;\widehat{\eeta}_{\tau,n},\widehat{t}_n|\bW_{ij})-Q_Y(\tau-\Delta_n;\widehat{\eeta}_{\tau,n},\widehat{t}_n|\bW_{ij})}\right\}
$$
where $\Delta_n$ is the bandwidth. We follow \cite{hall1988distribution} and  choose
$$
\Delta_n=1.57n^{-1/3}(1.5\phi^2\{\Phi^{-1}(\tau)\}/[2\{\Phi^{-1}(\tau)\}^2+1])^{1/3}
$$
 where $\Phi(\cdot)$ and $\phi(\cdot)$ are the distribution and density function for standard normal distribution. In addition, $\widehat{\LLambda}_{n12}=(\widehat{\llambda}_{n12,1}^\top,\cdots,
\widehat{\llambda}_{n12,K})^\top$ is a $K(q+3)\times1$ vector with $k$th element
$$
\widehat{\llambda}_{n12,k}=n^{-1}\sum_{i=1}^N\sum_{j=1}^{n_i}\hat{f}_{ij}
\{Q_Y(\tau_k;\widehat{\ttheta}_n|\bW_{ij})\}\{\widehat{\beta}_{n,1,\tau_k}I(
X_{ij}\leq\widehat{t}_n)+\widehat{\beta}_{n,2,\tau_k}I(X_{ij}>\widehat{t}_n)\}\bX_{ij}(\widehat{t}_n).
$$
 $\widehat{\LLambda}_{n22}$ is a scalar whose expression is $-n^{-1}\sum_{i=1}^N\sum_{j=1}^{n_i}\hat{f}_{ij}\{
Q_Y(\tau_k;\widehat{\ttheta}_n|\bW_{ij})\}\{\widehat{\beta}_{n,1,\tau_k}^2I(
X_{ij}\leq\widehat{t}_{n})+\widehat{\beta}_{n,2,\tau_k}^2I(X_{ij}>\widehat{t}_n)\}$.

A consistent estimator for $\bH$ is
\begin{eqnarray}
\widehat{\bf H}_n=n^{-1}\sum_{k=1}^K\sum_{l=1}^K\left\{\begin{array}{ll}
\left[\sum_{i,j}\tau_k(1-\tau_k)
{\bf h}_k({\bf W}_{ij};\widehat{\ttheta}_n){\bf h}_k({\bf W}_{ij};\widehat{\ttheta}_n)^\top\right. & \\
\left.+\sum_i\sum_{j\neq j^{'}}\left(\widehat{\delta}_{ijj^{'}}^{(\tau_k)}-\tau_k^2\right){\bf h}_k({\bf W}_{ij};\widehat{\ttheta}_n)
{\bf h}_k({\bf W}_{ij^{'}};\widehat{\ttheta}_n)^\top\right] & (k=l)\\
&\\
\left[\sum_{i,j}(\tau_k\wedge\tau_l-\tau_k\tau_l)
{\bf h}_k({\bf W}_{ij};\widehat{\ttheta}_n){\bf h}_l({\bf W}_{ij};\widehat{\ttheta}_n)^\top \right.& \\
\left.+\sum_i\sum_{j\neq j^{'}}\left(\widehat{\xi}_{ijj^{'}}^{(\tau_k)}-\tau_k^2\right){\bf h}_k({\bf W}_{ij};\widehat{\ttheta}_n)
{\bf h}_l({\bf W}_{ij^{'}};\widehat{\ttheta}_n)^\top\right] & (k\neq l).\\
\end{array}
\right.
\end{eqnarray}
One difficulty here is how to estimate $\widehat{\delta}_{ijj^{'}}^{(\tau_k)}$ and $\widehat{\xi}_{ijj^{'}}^{(\tau_k)}$ since it depends on the  correlated structure within each individual. \cite{li2015quantile} provided four kinds of structures, they are compound symmetry, AR(1) structure, heteroscedastic correlation  and unstructured correlation. We directly adopt the method of \cite{li2015quantile} to estimate $\widehat{\delta}_{ijj^{'}}^{(\tau_k)}$ and $\widehat{\xi}_{ijj^{'}}^{(\tau_k)}$ and omit the detailed computations  for saving space.

\section{Inference for Kink Point}
\subsection{Test for the existence of kink effect}
Above parameters estimation  and construction of interval are meaningful if and only if the change point significantly exists for each $\tau_k$, $k=1,\cdots,K$.   So how to statistically test for the existence of change point for each quantile level deserves to be explored. For any quantile level $\tau\in\{\tau_1,\cdots,\tau_K\}$, \cite{li2015quantile} defined the objective function
\begin{equation}\label{eq:ss}
L_n(\eeta_\tau,t)=n^{-1}\sum_{i=1}^N\sum_{j=1}^{n_i}\rho_\tau\{Y_{ij}-\bX_{ij}(t)\trans\eeta_\tau\}.
\end{equation}
We are interested in the following null ($H_0$) and alternative ($H_1$) hypothesis
\begin{equation}\label{eq:4.1}
\text{$H_0$: $\beta_{1,\tau}=\beta_{2,\tau}$ for any $t\in\mathcal{T}$\quad v.s.\quad $H_1:$   $\beta_{1,\tau}\neq\beta_{2,\tau}$  for some $t\in\mathcal{T}$,}
\end{equation}
where $\mathcal{T}$ is a compact set for $t$. Under the null hypothesis, the objective function becomes
$$
\widetilde{L}_n=n^{-1}\sum_{i=1}^N\sum_{j=1}^{n_i}\rho_\tau(Y_{ij}-\widetilde{\bX}_{ij}\trans\zzeta_\tau)
$$
where $\widetilde{\bX}_{ij}=(1,X_{ij},\bZ_{ij}\trans)\trans$ and $\zzeta_\tau=(\alpha_\tau,\beta_{1,\tau},\ggamma_\tau\trans)\trans$. In fact, $\widetilde{L}_n=\arg\min_{\beta_{1,\tau}=\beta_{2,\tau}}L_n(\eeta_\tau,t)$.
 In this paper, we proposed a sup-likelihood-ratio (SLR) test for testing the existence of change point. The SLR statistics is defined as
\begin{equation}
SLR_n(\tau)=\sup_{t\in\mathcal{T}}n\left\{\widetilde{L}_n-L_n(\widehat{\eeta}_n,\widehat{t}_n)\right\}
\end{equation}

 To investigate the asymptotic properties of proposed SLR test statistic, we consider the following local  alternative model $H_n$
\begin{eqnarray}\nonumber
%Q_Y(\tau;\eeta_\tau,t|\bW_{ij})&=&\alpha_\tau+\beta_{1,\tau}(X_{ij}-t)I(X_{ij}\leq t)+(\beta_{1,\tau}+n^{-1/2}\beta_{\tau}^*)\\\label{eq:local}
%&&\times(X_{ij}-t)I(X_{ij}>t)+\bZ_{ij}\trans\ggamma_{\tau}
%Q_Y(\tau;\eeta_\tau,t|\bW_{ij})&=&\alpha_\tau+\beta_{1,\tau}X_{ij}+n^{-1/2}\Delta\beta_\tau(X_{ij}-t)I(X_{ij}>t)+\bZ_{ij}\trans\ggamma_{\tau}.
Q_Y(\tau;\eeta_\tau,t|\bW_{ij})&=&\alpha_\tau+\beta_{1,\tau}(X_{ij}-t)I(X_{ij}\leq t)+(\beta_{1,\tau}+n^{-1/2}\Delta\beta_{\tau})\\\label{eq:local}
&&\times(X_{ij}-t)I(X_{ij}>t)+\bZ_{ij}\trans\ggamma_{\tau}
\end{eqnarray}
where $\Delta\beta_\tau=\beta_{2,\tau}-\beta_{1,\tau}$. The following limiting results hold.

\begin{thm}\label{thm:slr1}
%\begin{itemize}
  Under the Assumptions (A1)-(A3) and the null hypothesis $H_0$, $SLR_n(\tau) \Rightarrow(1/2)\{\sup_{t\in\mathcal{T}}\mathcal{G}(t)^\top{ \mathcal{V}}(t)^{-1}\mathcal{G}(t)
-\mathcal{G}_1^\top\mathcal{V}_1^{-1}\mathcal{G}_1\}$ in distribution as $n\rightarrow\infty$, where $\mathcal{G}(t)$ is a mean-zero Gaussian process with covariance function
$$
{\bC}(t_1,t_2)=n^{-1}\left\{\sum_{i,j}{\bX}_{ij}(t_1){\bX}_{ij}(t_2)
\tau(1-\tau)+\sum_i\sum_{j\neq j^{'}}{\bX}_{ij}(t_1){\bX}_{ij^{'}}(t_2)\left(\delta_{ijj^{'}}^{(\tau)}
-\tau^2\right)\right\},
$$
and $\mathcal{V}(t)=n^{-1}\sum_{i,j}E\left\{{\bX}_{ij}(t){\bX}_{ij}(t)^\top
f_{ij}\left(\widetilde{\bX}_{ij}^\top\zzeta_{0,\tau}\right)\right\}$. $\mathcal{G}_1$ is also a mean-zero Gaussian process under $H_0$ whose covariance function is
$$
\widetilde{\bC}=n^{-1}\left\{ \sum_{i,j}\widetilde{\bX}_{ij}\widetilde{\bX}_{ij}^\top \tau(1-\tau)+\sum_i\sum_{j\neq j^{'}}\widetilde{\bX}_{ij}\widetilde{\bX}_{ij^{'}}^\top\left(\widetilde{\delta}_{ijj^{'}}^{(\tau)}
-\tau^2\right)\right\}.
$$
%\end{itemize}
\end{thm}

\begin{thm}\label{thm:slr2}
 Under the Assumptions (A1)-(A3) and the local alternative model $H_n$, as $n\rightarrow\infty$, we have
$$
SLR_n(\tau)\Rightarrow1/2\left[\sup_{t\in\mathcal{T}}\left\{
\mathcal{G}(t)+{\bP}(t)\right\}^\top\mathcal{V}(t)^{-1}\left\{\mathcal{G}(t)
+{\bP}(t)\right\}-(\mathcal{G}_1+\widetilde{\bP}_1)^\top\mathcal{V}_1^{-1}(
\mathcal{G}_1+\widetilde{\bP}_1)\right]
$$
where ${\bP}(t)=n^{-1}\sum_{i,j}\bX_{ij}(t)\Delta\beta_\tau(X_{ij}-t)I(X_{ij}>t)f_{ij}(\widetilde{\bX}_{ij}^\top
\zzeta_{0,\tau})$ and  $\widetilde{\bP}_1
=n^{-1}\sum_{i,j}\widetilde{\bX}_{ij}\Delta\beta_\tau(X_{ij}-t)I(X_{ij}>t)f_{ij}(\widetilde{\bX}_{ij}^\top
\zzeta_{0,\tau})$.
\end{thm}
From Theorems (\ref{thm:slr1}) and (\ref{thm:slr2}), if $\Delta\beta_\tau\neq0$ i.e $H_1$ holds, ${\bP}(t)\neq0$ and $\widetilde{\bP}_1\neq0$. Then  SLR test statistic would converge to a different limiting distribution from that under $H_0$. ${\bP}(t)$ and $\widetilde{\bP}_1$ here serve to distinguish the null hypothesis from the alternative hypothesis. Since the null distribution of $SLR_n(\tau)$ takes nonstandard form, its critical values cannot be tabulated directly. To generate the critical values, we  propose a blockwise  wild bootstrap method to characterize the limiting behavior of $SLR_n(\tau)$ under $H_0$. Different from the wild bootstrap method in \cite{lee2011testing} and \cite{zhang2014testing}, we treat the observations within a subject as a block and draw disturbing sample only for the $N$ subjects, so-called blockwise bootstrap. The procedures go as follows.

%SLR test statistic converges to different limiting distribution under $H_0$ and $H_1$ depending on ${\bP}(t)$ and $\widetilde{\bP}_1$.

\begin{algorithm}[!h]
\caption{~~Calculate the P-values for SLR test statistic }\label{alg:4}
\begin{algorithmic}
\STATE {\bf Step 1.} Generate iid standard normal variable $\{u^{(b)}_{i};i=1,\cdots,N\}$ for $b=1,\cdots,B$ where $B$ is a sufficiently large  positive integer.
\STATE {\bf Step 2.} Calculate the following functions, respectively
\begin{align*}
\mathcal{G}^{(b)}(t)&=n^{-1/2}\sum_{i=1}^Nu_i^{(b)}\sum_{j=1}^{n_i}\bX_{ij}(t)\left[\tau-I\left\{
Y_{ij}-{\bX}_{ij}(t)^\top\widehat{\eeta}_{\tau}<0\right\}\right];\\
{\mathcal{G}}_1^{(b)}&=n^{-1/2}\sum_{i=1}^Nu_{i}^{(b)}\sum_{j=1}^{n_i}\widetilde{\bX}_{ij}\left\{\tau-I\left(
Y_{ij}-\widetilde{\bX}_{ij}^\top\widehat{\zzeta}_{\tau}<0\right)\right\}.
\end{align*}
\STATE{\bf Step 3.}  Calculate the test statistic $\{SLR_n^{(b)}(\tau)\}_{b=1}^B$:
$$
SLR_n^{(b)}(\tau)=1/2\sup_{t\in\mathcal{T}}\left\{\mathcal{G}^{(b)}(t)^\top\widehat{\mathcal{V}}
^{-1}\mathcal{G}^{(b)}(t)-{\mathcal{G}}_1^{(b)\top}\widehat{\mathcal{V}}_1{\mathcal{G}}_1^{(b)}\right\}
$$
where
\begin{align*}
\widehat{\mathcal{V}}(t)&=n^{-1}\sum_{i,j}\bX_{ij}(t)\bX_{ij}(t)^\top \hat{f}_{ij}\{ \bX_{ij}(t)^\top\widehat{\eeta}_\tau\}\\
\widehat{\mathcal{V}}_1&=n^{-1}\sum_{i,j}\widetilde{\bX}_{ij}\widetilde{\bX}_{ij}^\top \hat{f}_{ij}(\widetilde{\bX}_{ij}^\top\widehat{\zzeta}_{\tau})
\end{align*}
\STATE{\bf Step 4.} The asymptotic P-value is
$$
\widehat{P}_n=1/B\sum_{b=1}^BI\left\{SLR_n^{(b)}(\tau)>SLR_n(\tau)\right\}
$$
\end{algorithmic}
\end{algorithm}

\subsection{Test-inversion confidence set for kink point}
%As found by Li (2015), the coverage rate for Wald-type confidence interval for the change point is always lower than the nominal level 95\%. As a matter of fact, such phenemenon extensively exists in this type of continuous threshold regression see for example Hansen (2017) and Zhang (2017). Fong (2018) proposed an test-inversion confidence set for change point to  improve the coverage.
In this subsection, we propose three types of confidence intervals (CI) for the common change point. First and foremost, the Wald-type CI can be directly constructed based on the asymptotic normality in Theorem \ref{thm1} i.e. $\widehat{t}\pm z_{\alpha/2}\text{SE}(\widehat{t})$ where $z_{\alpha/2}$ is the $\alpha/2$th upper quantile of the standard normal  and $\text{SE}(\widehat{t})$ is the standard error of $\widehat{t}$ obtained by estimating $\widehat{\SSigma}_n$. Secondly, the bootstrap resampling is another popular method to construct the CI. The literatures on resampling methods in quantile regression is vast but in  longitudinal data, the extensively used  method is the subject bootstrap. Specially, we draw data from the original subject level triples $\{(Y_{ij},X_{ij},\bZ_{ij}):j=1,\cdots,n_i\}$ randomly with replacement for $B$ times. The bootstrap CI is defined as the $(\alpha/2)$th and $(1-\alpha/2)$th quantiles of the bootstrap estimators $\{\widehat{t}^*_b,b=1,\cdots,B\}$.

The third type of CI is constructed by inversion a proposed test statistics for finding a set of null values that is not rejected at pre-specified confidence level.  Therefore, we are interested in the following hypotheses
\begin{equation}\label{eq:CI}
\text{$H_0:$ $t_{\tau_k}=t_0$ for all $k=1,\cdots,K$ v.s. $H_1$: $t_{\tau_k}\neq t_0$ for some $k$},
\end{equation}
where $t_0$ is a  candidate change point. The null hypothesis implies that the change points at all $K$ quantiles share a common value $t_0$, which exhibits homogeneity for $t$.

We build a rank score test  statistic for (\ref{eq:CI}). Under $H_0$, the regression coefficients $\eeta_{\tau_k}$ can be obtained by fitting the standard linear quantile regression with $t_{\tau_k}=t_0$. The resulting estimators are denoted as $\widehat{\eeta}_{\tau_k}(t_0)$  and the corresponding residuals  $\hat{u}_{ij,\tau_k}=Y_{ij}-Q_Y\{\tau_k;\widehat{\eeta}_{\tau_k}(t_0)|\bW_{ij}\}$ for $k=1,\cdots,K$. Then, the first order derivative of $Q_Y\{\tau_k;{\eeta}_{\tau_k},t|\bW_{ij}\}$ w.r.t parameter $t$ evaluated at $t=t_0$ and ${\eeta}_{\tau_k}={\eeta}_{\tau_k}(t_0)$ is $b_{ij}\{\tau_k;{\eeta}_{\tau_k}(t_0),t_0\}=-\beta_{1,\tau_k}I(X_{ij}\leq t_0)-\beta_{2,\tau_k}I(X_{ij}>t_0)$.

 We define the rank score test statistic as
\begin{equation}\label{eq:RS}
RS_n=\bT_n\trans\PPsi_n^{-1}\bT_n
\end{equation}
where $\bT_n=(T_{n,1},\cdots,T_{n,K})\trans$ is a $K\times1$ vector with $T_{n,k}=n^{-1/2}\sum_{i=1}^N\sum_{j=1}^{n_i}b_{ij}^*\{\tau_k;\widehat{\eeta}_{\tau_k}(t_0),t_0\}\psi_{\tau_k}(\hat{u}_{ij,\tau_k})$ and $\PPsi_n$ is a $K\times K$  matrix with $(k, l)$th   element for  $k,l=1,\cdots,K$  denoted as $\Psi_{n,k,l}$,
\begin{eqnarray}
\Psi_{n,k,l}=n^{-1}\left\{\begin{array}{ll}
\left[\sum_{i=1}^N\sum_{j=1}^{n_i}\tau_k(1-\tau_k)
b_{ij}^*\{\tau_{k};\widehat{\eeta}_{\tau_{k}}(t_0),t_0\}b_{ij}^*\{\tau_{l};\widehat{\eeta}_{\tau_{l}}(t_0),t_0\}+\right. & \\
\left.\sum_i\sum_{j\neq j^{'}}\left(\delta_{ijj^{'}}^{(\tau_k)}-\tau_k^2\right)b_{ij}^*\{\tau_{k};\widehat{\eeta}_{\tau_{k}}(t_0),t_0\}b_{ij^{'}}^*\{\tau_{l};\widehat{\eeta}_{\tau_{l}}(t_0),t_0\}\right] & (k=l)\\
&\\
\left[\sum_{i=1}^{N}\sum_{j=1}^{n_i}(\tau_k\wedge\tau_l-\tau_k\tau_l)
b_{ij}^*\{\tau_{k};\widehat{\eeta}_{\tau_{k}}(t_0),t_0\}b_{ij}^*\{\tau_{l};\widehat{\eeta}_{\tau_{l}}(t_0),t_0\}+\right. & \\
\left.\sum_i\sum_{j\neq j^{'}}\left(\xi_{ijj^{'}}^{(\tau_k)}-\tau_k\tau_l\right)b_{ij}^*\{\tau_{k};\widehat{\eeta}_{\tau_{k}}(t_0),t_0\}b_{ij^{'}}^*\{\tau_{l};\widehat{\eeta}_{\tau_{l}}(t_0),t_0\}\right] & (k\neq l)\\
\end{array}
\right.
\end{eqnarray}
 The $b_{ij}^*\{\tau_{k};\widehat{\eeta}_{\tau_{k}}(t_0),t_0\}$ is defined as follows. Let $\bM(t)=(\bX_{11}(t),\bX_{12}(t),\cdots,\bX_{N,n_N}(t))\trans$ be $n\times(q+3)$ matrix and $\bB_k=(b_{11}\{\tau_k;\widehat{\eeta}_{\tau_k}(t_0),t_0\},b_{12}\{\tau_k;\widehat{\eeta}_{\tau_k}(t_0),t_0\},\cdots,b_{N,n_N}\{\tau_k;\widehat{\eeta}_{\tau_k}(t_0),t_0\})\trans$ be a $n\times1$ vector. Furthermore, define $\bB_k^*\equiv(\bI_n-\bV_k)\bB_k$ where $\bI_n$ is $n\times n$ identity matrix,  $\bV_k=\bM(t_0)\{\bM(t_0)\trans\UUpsilon_k\bM(t_0)\}^{-1}\bM(t_0)\trans\UUpsilon_k$ and $\UUpsilon_k$ is a $n\times n$ diagonal matrix with elements $\hat{f}_{11}\{Q_Y(\tau_k;\eeta_{\tau_k,0},t_0|\bW_{ij})\}$ for $i=1,\cdots,N$ and $j=1,\cdots,n_i$. So, $\bB_k^*$ is actually the projection of partial score vector $\bB_k$ on $\bM(t_0)$.

 % $n^{-1}\sum_{i=1}^N\sum_{j=1}^{n_i}b_{ij}^*\{\tau_{k_1};\widehat{\eeta}_{\tau_{k_1}}(t_0),t_0\}$ $b_{ij}^*\{\tau_{k_2};\widehat{\eeta}_{\tau_{k_2}}(t_0),t_0\}\psi_{\tau_{k_1}}(\hat{u}_{ij,\tau_{k_1}})\psi_{\tau_{k_2}}(\hat{u}_{ij,\tau_{k_2}})$.

We assume the following conditions to study the asymptotic property of $RS_n$.
\begin{itemize}
\item[(A6)] The Lebesgue density $f_{ij}^{\tau_k}(\cdot)$ has a bounded first-order derivative for all $i,j$ and $k$.
\item[(A7)] The smallest eigenvalue of $\PPsi_n$ is bounded away from zero as $n\rightarrow\infty$.
\end{itemize}
Assumption (A6) is an important condition in deriving the limiting behavior of rank score statistic and Assumption (A7) requires that the matrix $\PPsi_n$ is strictly positive definite. Both the two conditions can also be found in \cite{zhang2017composite}.
\begin{thm}\label{thm:RS}
Under the Assumptions (A1)-(A4) and (A6)-(A7),  and the null hypothesis $H_0$ in (\ref{eq:CI}), we have $RS_n\stackrel{d}{\longrightarrow}\chi^2_{K}$, as $n\rightarrow\infty$.
\end{thm}

Based on Theorem \ref{thm:RS}, we develop a rank score test inversion set for the kink point  and the detailed steps can be found in Algorithm \ref{alg:2}.

\begin{algorithm}
\caption{~~Rank score test inversion CI for change point}\label{alg:2}
\begin{algorithmic}
\STATE {\bf Step 1.}  Estimate $\widehat{t}$ by using the profiled estimation procedures in section \ref{sec:21}.
\STATE {\bf Step 2.} Test $H_0:t=\widehat{t}+k\delta$ for $k=1,2,\cdots,$ at the significance level of 0.05 by using the proposed test statistic (\ref{eq:RS}) where $\delta$ is a small positive increment. The upper bound is the minimum accepted point $\widehat{t}^U=\widehat{t}+U\delta$ for $k=U$.
\STATE {\bf Step 3.} Following the similar procedure in Step 2, we can   search for the lower bound $\widehat{t}^L$ as the maximum accepted point. The 95\% rank score test inversion CI  is $[\widehat{t}^U,\widehat{t}^L]$.
\end{algorithmic}
\end{algorithm}

\begin{rem}
In the special case where the error term is homoscedastic, that is, $f_{ij}^{(\tau_k)}(\cdot)=f^{(\tau_k)}(\cdot)$ for all $i$, $j$ and $k$, then
$$
\bV_k=\bM(t_0)\{\bM(t_0)\trans\bM(t_0)\}^{-1}\bM(t_0),
$$
and the quantile rank score test does not require estimating the density $f_{ij}^{(\tau)}(\cdot)$.
\end{rem}

%%%%%%%%%%%%%%%%%%%%%%%%%%%%%%%%%%%%%%%%%%%%%%%%%%%%%%%%%%%%%%%%%%%%%%%%%%%%%%%%%%%%%%%%%%%%%%%%%%%%%%%%%%%%%%%%%%%%%%%%%%%%

\section{Simulation Studies}
\subsection{Setup}
In this section, we study  the finite sample performance for the proposed methods. The simulation data were generated from the following setting
\begin{equation}\label{eq:sim}
Y_{ij}=\alpha+\beta_1(X_{ij}-t)I(X_{ij}\leq t)+\beta_2(X_{ij}-t)I(X_{ij}>t)-\gamma Z_{ij}+e_{ij}.
\end{equation}
where $t=5$ is the change point, $(\alpha,\beta_1,\beta_2,\gamma)\trans=(3,1,-1,0.2)\trans$ are regression coefficients and $e_{ij}$ is error term.  Four different cases are considered:

{\bf Case 1.} A random effect  model with $e_{ij}=a_i+\epsilon_{ij}$ where $a_i\stackrel{iid}{\sim}N(0,1)$ and $\epsilon_{ij}\stackrel{iid}{\sim}N(0,1)$.

{\bf Case 2.} An  AR(1) correlation model with  $e_{ij}=v(X_{ij})u_{ij}$ where  $v(X_{ij})=3.2-0.2X_{ij}$, $u_{ij}=0.5u_{i,j-1}+\epsilon_{ij}$ and $\epsilon_{ij}\sim N(0,1)$.

{\bf Case 3.} A heteroscedastic correlation model with $e_{ij}=a_i+g(X_{ij})\epsilon_{ij}$ where $a_i\stackrel{iid}{\sim}N(0,1)$ and  $g(X_{ij})=\sqrt{(3.2-0.2X_{ij})^2-1}$.

{\bf Case 4.} A  random effect  model with $e_{ij}=a_i+\epsilon_{ij}$ where $a_i\stackrel{iid}{\sim}N(0,1)$ and $\epsilon_{ij}\stackrel{iid}{\sim}t_3$.

Cases 1-3 are similar to that of \cite{li2015quantile} and Case 4 considers the heay-tailed error. In Cases 1, 3 and 4, $X_{ij}\stackrel{iid}{\sim}U(0,10)$ and in Case2, the threshold variable was generated from  $X_{ij}=X_{i,j-1}+0.5$ for $j>1$ and $X_{i1}\stackrel{iid}{\sim}U(0.5,7.5)$. For all cases, we let  $Z_{ij}\stackrel{iid}{\sim}U(0,10)$. The number of individuals is set to be $N=200$ and 400. To add  imbalance for the number of subjects, we let the number of observations $n_i=5$ for  $i=1,\cdots,N-2$ and $n_{N-1}=4$ for $(N-1)$th individual and $n_N=6$ for $N$th individual. Therefore, there are totally 1000 and 2000 observations respectively. For each scenario, we conduct 500 simulations.

%We consider three different  correlation structure as in Li (2015) and totally  six different cases. Cases 1-3 were generated from compound symmetry correlation structure where $X_{ij}\stackrel{iid}{\sim}U(0,10)$ and $e_{ij}=a_i+\epsilon_{ij}$ with $a_i\stackrel{iid}{\sim}N(0,1)$, $\epsilon_{ij}\stackrel{iid}{\sim}N(0,1)$ and $t_3$ corresponding to Case 1 and 2 respectively. In Case 3, we let $t=3$ for asymmetric change point. Cases 4-5 were generated from heteroscedastic correlation structure where $X_{ij}\stackrel{iid}{\sim}U(0,10)$ and $e_{ij}=a_i+g(X_{ij})\epsilon_{ij}$ with $a_i\stackrel{iid}{\sim}N(0,1)$ and  $g(X_{ij})=\sqrt{(3.2-0.2X_{ij})^2-1}$. The error term were generated from  $\epsilon_{ij}\stackrel{iid}{\sim}N(0,1)$  for Case 4 and  $\epsilon_{ij}\stackrel{iid}{\sim}t_3$ for Case 5. Case 6 was generated from AR(1) correlation structure where $e_{ij}=v(X_{ij})u_{ij}$, $v(X_{ij})=3.2-0.2X_{ij}$, $u_{ij}=0.5u_{i,j-1}+\epsilon_{ij}$ and $\epsilon_{ij}\sim N(0,1)$. The threshold variable was set as $X_{ij}=X_{i,j-1}+0.5$ for $j>1$ with $X_{i1}\stackrel{iid}{\sim}U(0.5,7.5)$. For all cases, we let  $Z_{ij}\stackrel{iid}{\sim}U(0,10)$. The number of individuals is set to be $N=200$ and 400. To add some inbalance for the number of subjects, we let the number of observations $n_i=5$ for individuals $i=1,\cdots,N-2$ and $n_{N-1}=4$ for $(N-1)$th individual and $n_N=6$ for $N$th individual. Therefore, there are totally 1000 and 2000 observations respectively. For each scenario, we conduct 500 simulations.

\subsection{Parameters estimation}
We first evaluate the sample performance of the proposed composite quantile regression (CQR) estimator. The quantile indices are set as $\tau_k=k/10$  for $k=3,\cdots,7$.  For comparison, we take two kinds of estimators into consideration. One is the  least absolute deviation (LAD) estimator proposed by \cite{li2015quantile}  and can be implemented by using the R code available at \url{https://onlinelibrary.wiley.com/doi/10.1111/biom.12313}, the other is the least square (LS) estimator, which is a longitudinal version of \cite{hansen2017regression} and its implementation can be found at \url{https://github.com/ChuangWAN1994/CQRCPM/blob/master/LSCPM.R}.

%Considered that no one has investigated the similar longitudinal setting in mean regression, we also derive the corresponding  asymptotic properties. The details can be found in the Supplementary materials. In fact, the proposed LS estimators is actually the longitudinal version of the reression kink model of Hansen (2017).

%For comparsion, we also consider the least squares (LS) estimator and the least absolute deviation (LAD) estimator. The LAD method was proposed by Li (2015) and can be implemented by using the R code avaiable at \url{https://onlinelibrary.wiley.com/doi/10.1111/biom.12313}.

Table \ref{tab:1} summarizes the average  bias, the Monte Carlo  standard deviations (SD), the estimated standard errors (ESE) and the empirical coverage probability (ECP) of 95\% Wald-type confidence intervals for  LAD, LS and CQR estimators of change point $t$. From the Table, all the biases are ignorable, indicating the estimated kink points of three methods are consistent.
 In addition, the SDs are quite close to  ESEs for all methods, which illustrates the asymptotical normality for the kink estimators. For all cases, we can find that CQR estimators have smaller biases and MSEs than LAD estimators, which  exhibits higher estimation efficiency. This confirms the finite sample advantages of CQR method gained by pooling information from multiple quantiles. In Case 1-3 with normal errors, the CQR and LAD estimators are comparable to that of LS estimation, but  in Case 4 with heavy-tailed error, the estimators based on quantile regression (LAD and CQR) perform better than LS estimator with relatively small biases and MSEs.  This phenomenon  reflects the robustness advantage of quantile regression to mean regression. The coverage probabilities of Wald-type confidence intervals  are generally smaller than the nominal level 95\%. Although as the sample size increases to $N=400$, the coverage probabilities improves slightly.  Such poor performance also appears in \cite{li2015quantile} and \cite{hansen2017regression} and  we will show  in section 4.4 that our proposed test-inversion set based on quantile rank score can help to improve the coverage probabilities of CQR estimator.

%We compare the proposed composite change point estimator with the least absolute deviation (LAD) estimator proposed by Li (2015) and the least square (LS) estimator. Hansen (2017) proposed a regression kink model based on mean regression for stationary weakly dependent data. For comparsion, we extend his model into  But our estimators are more close to the true parameter with relatively smaller biases than the other two types of estimators.

\begin{table}
\label{tab:1}
\centering
\begin{threeparttable}
\footnotesize
\caption{Simulation results of different kink point estimators. Average bias (Bias) is multiplied by a factor of 10. }
\begin{tabular}{cccccccc}
\hline
\hline
\multirow{2}{*}{Case} & & \multicolumn{3}{c}{$N=200$} & \multicolumn{3}{c}{$N=400$} \\
\cmidrule(lr){3-5}\cmidrule(lr){6-8}
 & &LAD & LS & CQR & LAD & LS & CQR \\
\hline
Case 1 & Bias & 0.009 & 0.013 & 0.0021 & -0.009 & -0.013 & 0.005 \\
       & SD & 0.118 & 0.096 & 0.106 & 0.084 & 0.065 & 0.072\\
       & ESE & 0.114 & 0.089 & 0.099 & 0.079 & 0.063 & 0.069 \\
       & MSE & 0.014 & 0.009 & 0.011 & 0.007 & 0.004 & 0.005 \\
       &ECP & 0.926 & 0.940 & 0.916 & 0.940 & 0.938 & 0.920 \\
\hline
Case 2 & Bias & -0.003 & -0.150 & -0.006 & 0.003 & 0.035 &-0.002\\
       & SD & 0.247 & 0.195 & 0.224 & 0.191 & 0.148 & 0.147 \\
       & ESE & 0.208 & 0.174 & 0.186 & 0.188 & 0.144 & 0.157 \\
       & MSE & 0.061 & 0.038 & 0.050 & 0.038 & 0.026 & 0.025 \\
       &ECP & 0.902 & 0.906 & 0.888 & 0.907 & 0.910 & 0.910 \\
\hline
Case 3 & Bias & 0.028 & -0.043 & -0.025 & -0.044 & -0.040 & -0.057 \\
       & SD & 0.196 & 0.158 & 0.173 & 0.134 & 0.107 & 0.115\\
       & ESE & 0.176 & 0.141 & 0.153 & 0.124 & 0.100 & 0.108 \\
       & MSE & 0.038 & 0.025 & 0.030 & 0.018 & 0.012 & 0.013 \\
       &ECP & 0.914 & 0.920 & 0.892 & 0.918 & 0.930 & 0.930 \\
\hline
Case 4 & Bias & -0.381 & -0.666 & 0.216 & 0.076 & 0.089 & 0.059 \\
       & SD & 0.520 & 0.709 & 0.443 & 0.094 & 0.096 & 0.085\\
       & ESE & 0.387 & 0.447 & 0.336 & 0.089 & 0.088 & 0.078 \\
       & MSE & 0.271 & 0.507 & 0.196 & 0.009 & 0.009 & 0.007 \\
       &ECP & 0.846 & 0.828 & 0.842 & 0.930 & 0.928 & 0.920 \\

\hline
\end{tabular}
\end{threeparttable}
\end{table}

\subsection{Power analysis}
\begin{figure}[!h]
\centering
\includegraphics[scale=0.9]{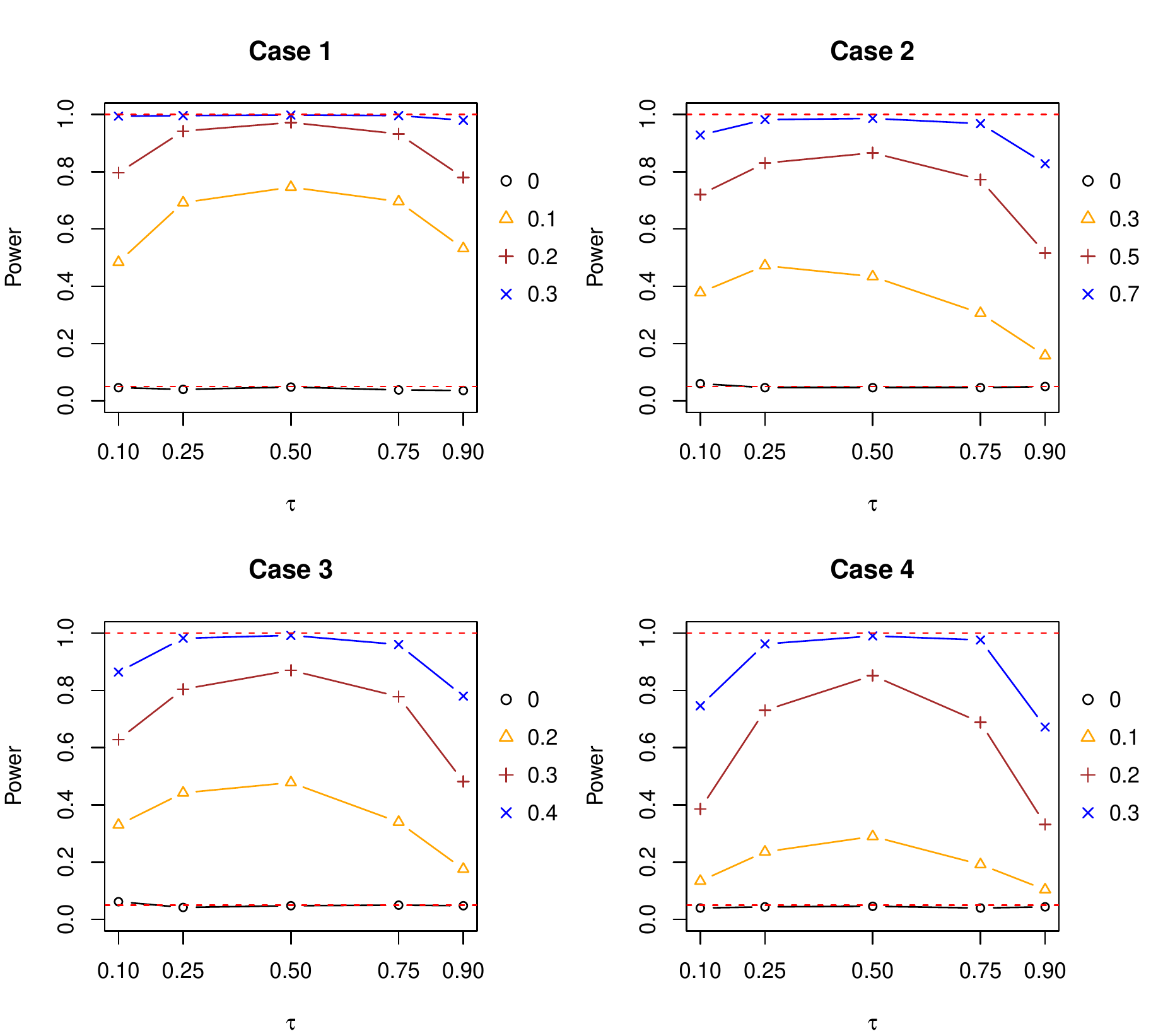}
%\vspace{-0.8cm}
\caption[]{\label{fig:1}
Power across different $\tau$'s for Case 1-4 with varying $\Delta\beta$ based on $N=200$ individuals.}
\end{figure}

 To evaluate the Type I error and local power of proposed test in Algorithm \ref{alg:4}, we conduct another simulation study with varying $\beta_2=\beta_1+\Delta\beta$ in  model (\ref{eq:sim}) where $\Delta\beta$ is from 0 to some values, and other parameters are kept as before. For each case, the P-values are obtained by 300 bootstrap replicates based on the sample size $n=1000$. The results are illustrated in Figure \ref{fig:1}. As shown in the Figure, when $\Delta\beta=0$ (the lines with black circles),  the powers of each case all around the nominal level 5\%, suggesting that our method has reasonable control of Type I errors.  As expected, as $\Delta\beta$ increases, i.e. the kink effects get strengthened,  the local power across different $\tau$'s all gradually approach one.  This suggests that our proposed test has decent power to detect the kink effects at different quantiles. We also observe that the powers at non-extreme quantiles such as $\tau=0.5$ are always better than extreme quantiles such as $\tau=0.1,0.9$. It is common in quantile test due to the asymmetry of observations at tail quantiles and can be improved with the sample size increases.

%The local powers obtained by 500 repetition based on $n=1000$ observation with $K=1,2$ kink effects across different $c$. The lines with blcak cycles and  orange crosses represent the proposed test and \emph{Indep} test, respectively.

\subsection{Confidence intervals}
Last, we evaluate the test-inversion confidence intervals  based on quantile rank score  (QRS) test by comparing it to the blockwise bootstrap (Boot) intervals described in Section 3.2 and the Wald-type (Wald) intrevals. The bootstrap times is set to be $400$.   The estimated mean lengths (EML), the empirical coverage probabilities (ECP) and the average running time (in seconds) based on $N=200$ and $400$ of all cases  are summarized in Table \ref{tab:2}.

There is no doubt that the Wald method gives worst confidence intervals  for both $N=200$ and 500 among the three  constructions. In finite samples, the ECP of QRS method are, in general, more close to the nominal level than that of Boot, but the former leads relatively wider EMLs. However, QRS method costs much less computing time compared with Boot method. So it provides a good balance between the improvement of confidence interval and computational efficiency.

\begin{table}
\label{tab:2}
\centering
\begin{threeparttable}
\footnotesize
\caption{Comparsion of three types of confidence intervals: Wald, Boot and QRS for $N=200$ and 400. The nominal level is 0.95. }
\begin{tabular}{cccccccccc}
\hline
\hline
\multirow{2}{*}{Case}  & \multicolumn{3}{c}{Wald} & \multicolumn{3}{c}{Boot} & \multicolumn{3}{c}{QRS} \\
\cmidrule(lr){2-4}\cmidrule(lr){5-7}\cmidrule(lr){8-10}
 & ECP & EML & Time(s) & ECP & EML & Time(s) & ECP & EML & Time(s)    \\
\hline
Case 1 & 0.916 & 0.445 & 7.060 & 0.942 & 0.419 & 357.590 & 0.958 & 0.591 & 10.580 \\
Case 2 & 0.888 & 0.816 & 7.960 & 0.944 & 0.958 & 430.820 & 0.970 & 1.362 & 14.250 \\
Case 3 & 0.892 & 0.690 & 7.830 & 0.928 & 0.690 & 400.140 & 0.960 & 1.006 & 14.500 \\
Case 4 & 0.924 & 0.500 & 7.780 & 0.938 & 0.486 & 380.740 & 0.950 & 0.698 & 12.030 \\
\hline
Case 1 & 0.930 & 0.311 & 13.240 & 0.948 & 0.385 & 687.150 & 0.954 & 0.401 & 29.400 \\
Case 2 & 0.888 & 0.729 & 18.670 & 0.940 & 0.808 & 710.130 & 0.948 & 0.996 & 34.690 \\
Case 3 & 0.930 & 0.485 & 15.300 & 0.938 & 0.463 & 780.390 & 0.946 & 0.661 & 26.950\\
Case 4 & 0.922 & 0.347 & 13.100 & 0.928 & 0.324 & 740.060 & 0.948 & 0.470 & 35.610 \\
\hline
\end{tabular}
\end{threeparttable}
\end{table}

\section{Analysis of  Blood Pressure and Body Mass Index}
%We illustrate the proposed method to analyze the a subset data from the  Nation Growth , Lung and  Health study (NGHS).

It is well known that blood pressure  is an important indicator for human's health.
In chronic epidemiology, high blood pressure may lead to kinds of health problems such as coronary heart disease and stroke, while  low blood pressure will cause a shortage of blood to the body's organs and then some symptoms such as  the dizziness, the limb movement disorder are appeared. One  important topic in public health field is to study the relationship between the blood pressure (BP) and body mass index (BMI). Previous literatures suggested that BMI shows positive association with  BP (\cite{he1994body}, \cite{tesfaye2007association}), but some researcher found that the linear models are not sufficient to capture the positive relationship between BMI and BP. For instances, \cite{kerry2005blood} showed that there presents an significant nonlinear effect between BMI and diastolic BP for young women. \cite{zhang2014testing} formally demonstrated the existence of quantile threshold effect of  BMI  on systolic BP by using  quantile score test statistic. Moreover, \cite{zhang2017composite} studied  the composite estimation for change point across different quantiles between BMI and  systolic BP by analyzing the data from  the National Health and Nutrition Examination Survey (NHAENES).

In this section, we analyze a BMI and systolic BP longitudinal data from  the  Nation Growth , Lung and  Health Study (NGHS), avaiable at the NIH BioLINCC site (\url{https://biolincc.nhlbi.nih.gov/}). The NGHS is a  multi-center population-based cohort study  conducted to evaluate the longitudinal changes  of  childhood cardiovascular risk factors for 1166 Caucasian and 1213 African American girls. We only draw a subset of the first 300 subjects at the ages from 9 to 19. After removing some missing values, there are totally 2455 observations.  Different from the previous analysis , we examine the impacts of BMI on BP by using the proposed  methods to account for the dependence within one subject. Three quantile indices sets are considered, including  lower quantiles (LQ) set  $\{ 0.27,0.28,0.29,\cdots,0.33\}$,   median quantiles set (MQ) $\{0.48,0.49,0.50,0.51,0.52\}$ and  high quantiles set $\{0.77,0.78,0.79,\cdots,0.83\}$.

We first examine the existence of BMI kink effect by employing the proposed SLR test procedures in Algorithm \ref{alg:4} in Section 3.2 at each quantile level. The results of P-values and estimated kink point estimators are reported in Table \ref{tab:ana}. From the table, we observe that all the P-values approach zeros, suggesting  significant kink effects at all quantiles.  The kink estimators are quite close within one indices set.
To further check the commonality of kink points, we consider the following hypotheses
\begin{equation}\label{eq:common}
\text{$H_0:$ $t_{\tau_1}=\cdots=t_{\tau_K}$\quad v.s.\quad $H_1$: $t_{\tau_{k}}\neq t_{\tau_{l}}$ for some $k\neq l$}.
\end{equation}
For testing (\ref{eq:common}), we can construct the Wald type statistic based on the asymptotic properties in \cite{li2015quantile}. The resulting P-values for LQ, MQ and HQ are  0.101, 0.075 and 0.987, respectively, confirming the statistical existence of common kink points at the significance level 5\%.  To capture poential kink effects, we have the following longitudinal quantile regression model at a given $\tau$
$$
Q_Y(\tau|X_{ij},Z_{ij})=\alpha_\tau+\beta_{1,\tau}(X_{ij}-t)I(X_{ij}\leq t)+\beta_{2,\tau}I(X_{ij}>t)+\gamma_\tau Z_{ij},
$$
where $X_{ij}$ and $Z_{ij}$ denote BMI and the age, respectively. $(\alpha_\tau,\beta_{1,\tau},\beta_{2,\tau},\gamma_{\tau})\trans$ are unknown regression coefficients varying with $\tau$ and $t$ is unknown change point that are common within one indices set. By using the two-step estimation method described in Section 2.1, we can obtain the coefficients estimators across different $\tau$'s and the  composite change point estimator. Table \ref{tab:res}  summarizes the estimation results and the different types of confidence intervals of LS, LAD and CQR methods. For CQR method, we only report the results of $\tau=0.3$ for LQ, $\tau=0.5$ for MQ and $\tau=0.8$ for HQ.

\begin{table}
\label{tab:ana}
\centering
\begin{threeparttable}
\footnotesize
\caption{P-values of sup-likelihood-ratio test and the kink point estimators for LQ, MQ and HQ.}
\begin{tabular}{ccccccccc}
\hline
\hline
 & LQ &  & & MQ & & & HQ & \\
\cmidrule(lr){1-3}\cmidrule(lr){4-6}\cmidrule(lr){7-9}
$\tau$ & SLR & $\widehat{t}_{\tau}$ & $\tau$ & SLR & $\widehat{t}_{\tau}$  & $\tau$ & SLR & $\widehat{t}_{\tau}$  \\
\hline
0.27 & 0.005 & 26.335  & 0.48 & 0.000 & 28.441 & 0.77 & 0.005 & 29.246 \\
0.28 & 0.000 & 28.087  & 0.49 & 0.000 & 28.350 & 0.78 & 0.000 & 29.069 \\
0.29 & 0.005 & 26.750  & 0.50 & 0.000 & 28.461 & 0.79 & 0.005 & 29.123 \\
0.30 & 0.000 & 26.044  & 0.51 & 0.000 & 28.414 & 0.80 & 0.015 & 29.027 \\
0.31 & 0.000 & 27.856  & 0.52 & 0.000 & 28.484 & 0.81 & 0.015 & 29.022 \\
0.32 & 0.000 & 27.857  &      &       &        & 0.82 & 0.020 & 29.027 \\
0.33 & 0.000 & 28.179  &      &       &        & 0.83 & 0.015 & 29.022 \\
 \hline
\end{tabular}
\end{threeparttable}
\end{table}

\begin{figure}[!h]
\centering
\includegraphics[scale=0.9]{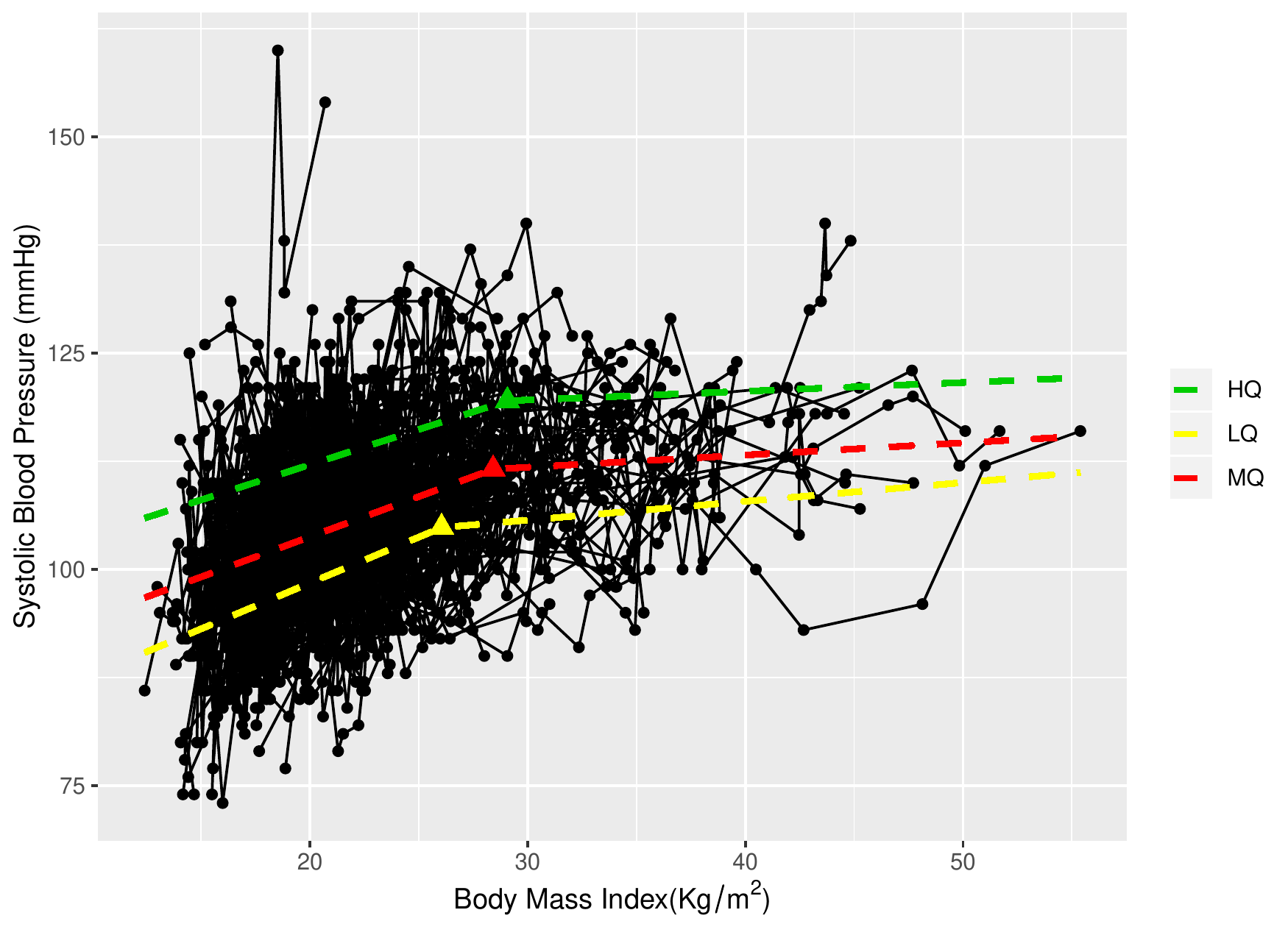}
%\vspace{-0.8cm}
\caption[]{\label{fig:bmi}
The fitted quantile   curves at different quantile levels for BMI against systolic BP. $\blacktriangle$ denotes the estimated kink points.}
\end{figure}

From the  Table \ref{tab:res}, the coefficients  show that  systolic BP firstly increases with BMI ($\beta_1>0$ of all methods), but with BMI reaching centain kink points,  the positive  growth relationship gets weaker ($\beta_2<\beta_1$). The estimated $\gamma$ all greater than zeros indicates a   positively effect of age on systolic BP.  This finding in accordance with  \cite{zhang2014testing} and \cite{zhang2017composite}. For different  methods, the kink point estimators are different. For LS estimators, it models the conditional mean of  systolic BP and the estimated kink point is around 26.225 kg/m$^2$. The LAD method is a  single quantile analysis  given by \cite{li2015quantile}. Interestingly, we find that as $\tau$  increases, the estimated kink points  increase from 26.004 kg/m$^2$ to 29.027 kg/m$^2$. Such phenomenon also appears in composite estimator whose kink estimators are 26.045 kg/m$^2$, 28.461 kg/m$^2$ and 29.069 kg/m$^2$ for LH, MQ and HQ respectively. This truth has a biological  intuition that people with high BP are more likely to possess  higher BMIs, therefore reaching the turning point later. Compared with LAD method, the proposed composite estimation gives shorter confidence intervals for the kink points, which indicates that combining information from multiple quantiles leads to more efficient estimation than only using a single quantile information. The fitted quantile curves of BMI against  systolic BP at LQ, MQ and HQ   in Figure \ref{fig:bmi} also illustrates our empirical findings.

\begin{sidewaystable}
%\begin{table}
\centering
\begin{threeparttable}
\label{tab:res}
\footnotesize
\caption{The estimated parameters, the standard errors (listed in parentheses) and the their confidence intervals from different estimating methods}
\begin{tabular}{cccccccc}
\hline
\hline
& \multirow{2}{*}{LS}  &  & LAD & & & CQR &   \\
\cmidrule(lr){3-5}\cmidrule(lr){6-8}
 &  & 0.3 & 0.5 & 0.8 & LQ & MQ & HQ \\
\hline
$\alpha$ & $\mathop{103.986}\limits_{(1.926)}$ & $\mathop{99.437}\limits_{(2.961)}$ & $\mathop{105.130}\limits_{(1.415)}$ & $\mathop{112.546}\limits_{(2.063)}$ & $\mathop{99.524}\limits_{(1.705)}$ & $\mathop{105.627}\limits_{(1.189)}$ & $\mathop{112.504}\limits_{(1.362)}$  \\
Wald & [100.211,107.760] & [93.621,105.253] & [102.357,107.903] & [108.497,116.595] & [96.182,102.865] & [102.850,107.510] & [109.835,115.174] \\
Boot & [98.019,108.054] & [93.248,103.672] & [97.768,110.080] & [106.958,120.005] & [94.214,103.791] & [98.359,109.340] & [105.697,119.221] \\
$\beta_1$ & $\mathop{0.971}\limits_{(0.121)}$ & $\mathop{1.059}\limits_{(0.013)}$ & $\mathop{0.924}\limits_{(0.004)}$ & $\mathop{0.815}\limits_{(0.017)}$ & $\mathop{1.060}\limits_{(0.116)}$ & $\mathop{0.928}\limits_{(0.066)}$ & $\mathop{0.814}\limits_{(0.127)}$  \\
Wald & [0.733,1.208] & [1.033,1.084] & [0.916,0.933] & [0.783,0.847] & [0.833,1.286] & [0.798,1.058] & [0.566,1.062] \\
Boot & [0.763,2.051] & [0.805,1.946] & [0.722,1.922] & [0.528,1.239] & [0.820,1.933] & [0.719,2.169] & [0.528,2.040] \\
$\beta_2$ & $\mathop{0.239}\limits_{(0.142)}$ & $\mathop{0.218}\limits_{(0.007)}$ & $\mathop{0.145}\limits_{(0.076)}$ & $\mathop{0.115}\limits_{(0.014)}$ & $\mathop{0.217}\limits_{(0.082)}$ & $\mathop{0.141}\limits_{(0.276)}$ & $\mathop{0.101}\limits_{(0.114)}$  \\
Wald & [-0.039,0.518] & [0.204,0.231] & [-0.005,0.294] & [0.088,0.142] & [0.057,0.378] & [-0.401,0.682] & [-0.122,0.324] \\
Boot & [-0.101,0.594] & [-0.047,0.579] & [-0.109,0.625] & [-0.849,2.834] & [-0.041,0.498] & [-0.225,0.562] & [-0.750,2.138] \\
$\gamma$ & $\mathop{0.421}\limits_{(0.088)}$ & $\mathop{0.446}\limits_{(0.006)}$ & $\mathop{0.433}\limits_{(0.005)}$ & $\mathop{0.404}\limits_{(0.009)}$ & $\mathop{0.441}\limits_{(0.076)}$ & $\mathop{0.429}\limits_{(0.073)}$ & $\mathop{0.410}\limits_{(0.093)}$  \\
Wald & [0.249,0.593] & [0.435,0.458] & [0.422,0.443] & [0.387,0.422] & [0.292,0.591] & [0.286,0.571] & [0.228,0.592] \\
Boot & [0.256,0.604] & [0.269,0.624] & [0.211,0.621] & [0.166,0.652] & [0.246,0.623] & [0.187,0.590] & [0.180,0.639] \\
$t$ & $\mathop{26.225}\limits_{(1.339)}$ & $\mathop{26.044}\limits_{(0.446)}$ & $\mathop{28.414}\limits_{(0.822)}$ & $\mathop{29.027}\limits_{(4.215)}$ & $\mathop{26.045}\limits_{(1.297)}$ &  $\mathop{28.461}\limits_{(0.625)}$& $\mathop{29.069}\limits_{(1.947)}$  \\
Wald & [23.601,28.849] & [22.891,29.197] & [26.803,30.024] & [20.767,37.288] & [23.502,28.587] & [27.236,29.687]  & [25.252,32.886] \\
Boot & [18.875,29.613] & [19.632,29.515] & [19.732,33.626] & [18.416,43.727] & [19.666,29.732] & [19.842,33.428] & [23.540,43.754] \\
Score &   &   &   &   & [19.040,29.936] & [25.456,31.372] & [23.227,38.416] \\

 \hline
\end{tabular}
%\begin{tablenotes}
   %    \footnotesize
  %     \item Wald: the Wald-type confidence intervals; Boot:
 %    \end{tablenotes}
\end{threeparttable}
%\end{table}
\end{sidewaystable}

\section{Discussion}
To aggregate the common kink point information from multiple quantiles, we proposed a new composite estimation method for kink quantile regression in longitudinal data.  Compared with the method in \cite{li2015quantile}, the proposed method can effectively capture the common kink effect. Both the simulation study and empirical analysis demonstrate that the composite estimating method is competitively efficient with the least square method and single quantile estimation method while more robust for heavy-tailed errors.

In this paper, to obtain composite estimator, we first find a index set including multiple quantiles and then verify its commonality. In reality, it is hard to find such quantile index set. Instead, it is more often that the neighbouring  quantiles shares the same kink point but at different regional quantile, the kink points are different. To solve this issue, a direct approach may adopt the regularization  method and the objective function becomes
$$
S_n(\ttheta)=n^{-1}\sum_{k=1}^K\sum_{i=1}^N\sum_{j=1}^{n_i}\rho_{\tau}\{Y_{ij}-Q_Y(\tau_k;\eeta_{\tau_k},t_{\tau_k}|\bW_{ij})\}+\sum_{k=2}^{K}p_{\lambda}(|t_{\tau_k}-t_{\tau_{k-1}}|)
$$
where $p_{\lambda}(\cdot)$ is some  penalty function such as LASSO \citep{tibshirani1996regression} and SCAD \citep{fan2001variable}. Reseach in this direction needs further investigation.

\scsection{APPENDIX}
\renewcommand{\theequation}{A.\arabic{equation}}
\renewcommand{\thesection}{A}
\vspace{-0.5cm}
\noindent
\subsection{Proof of Theorem \ref{thm1}:}
\begin{lem}\label{lem:consistency}
(Consistency) under Assumptions (A1)-(A3) and (A5),  $\widehat{\ttheta}$ is a strongly consistent estimator of $\ttheta_0$ as $n\rightarrow\infty$,
\end{lem}
\noindent
{\sc Proof of Lemma \ref{lem:consistency}:} The proof of this lemma is essentially the same as that in Theorem 1 of \cite{li2015quantile}. Note that for a fixed $t$, the objective function   is $S_n(\ttheta)$, which is equivalent to minimize
$$
n^{-1}\sum_{i=1}^N\sum_{j=1}^{n_i}\rho_{\tau_k}\{Y_{ij}-Q_Y(\tau_k;\eeta_{\tau_k}(t),t|\bW_{ij})\},
$$
for each $k=1,\cdots,K$. The rest of the proof shares the similar arguments in Theorem 1 of Li (2015). One can refer to their paper for more details and thus is omited  here.   $\hfill\blacksquare$ \\

\begin{lem}\label{lem:bahadur}
Define
\begin{align*}
u_i(\ttheta,\ttheta_0)=&\sum_{k=1}^K\sum_{j=1}^{n_i}\psi_{\tau_k}\{Y_{ij}-Q_Y(\tau_k;\eeta_{\tau_k},t|\bW_{ij})\}h_k(\bW_{ij};\ttheta)\\
&-\sum_{k=1}^K\sum_{j=1}^{n_i}\psi_{\tau_k}\{Y_{ij}-Q_Y(\tau_k;\eeta_{\tau_k,0},t_0|\bW_{ij})\}h_k(\bW_{ij};\ttheta_0)
\end{align*}
Suppose that the Assumptions (A1)-(A3) hold, then the following equation holds
\begin{equation}\label{eq:as}
\sup_{\|\ttheta-\ttheta_0\|\leq dn^{-1/2}}n^{-1/2}\Big\|\sum_{i=1}^N\left\{u_i(\ttheta,\ttheta_0)-E[u_i(\ttheta,\ttheta_0)]\right\}\Big\|=o_p(1)
\end{equation}
where $d$ is some positive constant.
\end{lem}
\noindent
{\sc Proof of Lemma \ref{lem:bahadur}:} By using the same argument in Theorem 2 of \cite{li2015quantile}, it is easy to show that
\begin{equation}\label{eq:bound}
\sum_{k=1}^K\sum_{i=1}^N\sum_{j=1}^{n_i}\psi_{\tau_k}\{Y_{ij}-Q_Y(\tau;\widehat{\eeta}_{\tau_k},\widehat{t}|\bW_{ij})\}h_k(\bW_{ij};\widehat{\ttheta})=o(n^{1/4}\left(\log\log n)^{1/2}\right)\quad\text{a.s.}
\end{equation}
Based on (\ref{eq:bound}),  Lemma  (\ref{lem:bahadur}) can be established directly by  Theorem 2.2 of \cite{he1996general} if the required conditions in that theorem hold.  Thus it is sufficient to verify the conditions (B1)-(B4) and (B5$^{'}$) of \cite{he1996general}.

For (B1), the measurability is directly satisfied.

For (B2), this can be  obtained  from the strong consistency in Lemma \ref{lem:consistency}.

To verify (B3),  we partition $u_i(\ttheta,\ttheta_0)$ based on the value of $X_{ij}$ such that
\begin{align*}
&U_i(\ttheta,\ttheta_0)\\
=&\sum_{k=1}^K\sum_{j=1}^{n_i}\Big[\psi_{\tau_k}\{Y_{ij}-Q_Y(\tau_k;\eeta_{\tau_k},t|\bW_{ij})\}h_k(\bW_{ij};\ttheta)-\psi_{\tau_k}\{Y_{ij}-Q_Y(\tau_k;\eeta_{\tau_k,0},t_0|\bW_{ij})\}h_k(\bW_{ij};\ttheta_0)  \Big]\\
&\times I\{X_{ij}>\max(t,t_0)\}+\sum_{k=1}^K\sum_{j=1}^{n_i}\Big[\psi_{\tau_k}\{Y_{ij}-Q_Y(\tau_k;\eeta_{\tau_k},t|\bW_{ij})\}h_k(\bW_{ij};\ttheta)-\psi_{\tau_k}\{Y_{ij}-\\
&Q_Y(\tau_k;\eeta_{\tau_k,0},t_0|\bW_{ij})\}h_k(\bW_{ij};\ttheta_0)  \Big]I\{t<X_{ij}\leq t_0\}+\sum_{k=1}^K\sum_{j=1}^{n_i}\Big\{\psi_{\tau_k}\{Y_{ij}-Q_Y(\tau_k;\eeta_{\tau_k},t|\bW_{ij})\}\\
&\times h_k(\bW_{ij};\ttheta)-\psi_{\tau_k}\{Y_{ij}-Q_Y(\tau_k;\eeta_{\tau_k,0},t_0|\bW_{ij})\}h_k(\bW_{ij};\ttheta_0)  \Big\}I\{t_0<X_{ij}\leq t\}+\sum_{k=1}^K\sum_{j=1}^{n_i}\Big[\\
&\psi_{\tau_k}\{Y_{ij}-Q_Y(\tau_k;\eeta_{\tau_k},t|\bW_{ij})\}h_k(\bW_{ij};\ttheta)-\psi_{\tau_k}\{Y_{ij}-Q_Y(\tau_k;\eeta_{\tau_k,0},t_0|\bW_{ij})\}h_k(\bW_{ij};\ttheta_0)  \Big]\\
&\times I\{X_{ij}\leq\min(t,t_0)\}\\
\equiv& U_{i1}(\ttheta,\ttheta_0)+U_{i2}(\ttheta,\ttheta_0)+U_{i3}(\ttheta,\ttheta_0)+U_{i4}(\ttheta,\ttheta_0).
\end{align*}
Then it is sufficient to show
$$
\sup_{\|\ttheta-\ttheta_0\|\leq dn^{-1/2}}\Big\|n^{-1/2}\sum_{i=1}^N\{U_{ij}(\ttheta,\ttheta_0)-E[U_{ij}(\ttheta,\ttheta_0)]\}\Big\|=o_p(1)
$$
for $j=1,2,3,4$.  For any $\|\ttheta-\ttheta_0\|\leq dn^{1/2}$, we have
\begin{align*}
&\|U_{i1}(\ttheta,\ttheta_0)\|\\
=&\Big\|\Big[\sum_{k=1}^K\sum_{j=1}^{n_i}\psi_{\tau_k}\{Y_{ij}-Q_Y(\tau_k;\eeta_{\tau_k},t|\bW_{ij})\}h_k(\bW_{ij};\ttheta)-\sum_{k=1}^K\sum_{j=1}^{n_i}\psi_{\tau_k}\{Y_{ij}-Q_Y(\tau_k;\eeta_{\tau_k,0},t_0|\bW_{ij})\}\\
&\times h_k(\bW_{ij};\ttheta_0)\Big] I\{X_{ij}>\max(t,t_0)\}\Big\|\\
\leq&\Big\|\sum_{k=1}^K\sum_{j=1}^{n_i}\psi_{\tau_k}\{Y_{ij}-Q_Y(\tau_k;\eeta_{\tau_k},t|\bW_{ij})\}\{h_k(\bW_{ij};\ttheta)-h_k(\bW_{ij};\ttheta_0)\}I\{X_{ij}>\max(t,t_0)\}\Big\|\\
&+\Big\| \sum_{k=1}^K\sum_{j=1}^{n_i}\left[\psi_{\tau_k}\{Y_{ij}-Q_Y(\tau_k;\eeta_{\tau_k},t|\bW_{ij})\}-\psi_{\tau_k}\{Y_{ij}-Q_Y(\tau_k;\eeta_{\tau_k,0},t_0|\bW_{ij})\}\right]h_k(\bW_{ij};\ttheta_0)\\
&\times I\{X_{ij}>\max(t,t_0)\}\Big\|\\
\equiv&\|I_{i1}\|+\|I_{i2}\|
\end{align*}
For $I_{i1}$, it is obvious that $E(\|I_{i1}\|^2|\bW_{ij})=o_p(1)$. For $I_{2i}$, we have
\begin{align*}
\|I_{i2}\|&\leq C_1\sum_{k=1}^K\sum_{j=1}^{n_i}\|\bU_{ij}\|I\{Q_1(\tau_k;\ttheta,\ttheta_0)\leq Y_{ij}\leq Q_2(\tau_k;\ttheta,\ttheta_0)\}I\{X_{ij}>\max(t,t_0)\}
\end{align*}
where $C_1$ is some positive constant, $\bU_{ij}=(1,X_{ij},\bZ_{ij}\trans)\trans$, $Q_1(\tau_k;\ttheta,\ttheta_0)$ and $Q_2(\tau_k;\ttheta,\ttheta_0)$ denote the  minimum and maximum values between $Q_Y(\tau_k;\eeta_{\tau_k},t|\bW_{ij})$ and $Q_Y(\tau_k;\eeta_{\tau_k,0},t_0|\bW_{ij})$. Thus
\begin{align*}
&E(\|I_{i2}\|^2|\bW_{ij})\\
\leq& C_1^2\sum_{j=1}^{n_i}\Big[\|\bU_{ij}\|^2I\{X_{ij}>\max(t,t_0)\}\Big]E\Big[\sum_{k=1}^K\sum_{l=1}^KI\{Q_1(\tau_k;\ttheta,\ttheta_0)\leq Y_{ij}\leq Q_2(\tau_k;\ttheta,\ttheta_0)\}\\
&\times I\{Q_1(\tau_l;\ttheta,\ttheta_0)\leq Y_{ij}\leq Q_2(\tau_l;\ttheta,\ttheta_0)\}\Big].
\end{align*}
Without loss of generality, we assume  $\max\{Q_1(\tau_k;\ttheta,\ttheta_0),Q_1(\tau_l;\ttheta,\ttheta_0)\}\leq\min\{Q_2(\tau_k;\ttheta,\ttheta_0),Q_1(\tau_l;\ttheta,\ttheta_0)\}$.
Let $Q_1(\tau_k,\tau_l)=\min\{Q_1(\tau_k;\ttheta,\ttheta_0), Q_1(\tau_l;\ttheta,\ttheta_0)\}$ and $Q_2(\tau_k,\tau_l)=\max\{Q_2(\tau_k;\ttheta,\ttheta_0), Q_2(\tau_l;\ttheta,\ttheta_0)\}$. We have
\begin{align*}
&E(\|I_{i2}\|^2|\bW_{ij})\\
\leq&C_1^2\sum_{j=1}^{n_i}\|\bU_{ij}\|^2\sum_{k=1}^K\sum_{l=1}^Kf_{ij}(\zeta_{kl})\{Q_2(\tau_k,\tau_l)-Q_1(\tau_k,\tau_l)\}I\{X_{ij}>\max(t,t_0)\}\\
\leq&C_2n^{-1/2}\sum_{j=1}^{n_i}\|\bU_{ij}\|^3\sum_{k=1}^K\sum_{l=1}^Kf_{ij}(\zeta_{kl})
\end{align*}
where the first inequality follows from the mean value theorem, $C_2$ is some positive constant, $\zeta_{kl}$ lies between $Q_1(\tau_k;\tau_l)$ and $Q_2(\tau_k,\tau_l)$. By Assumptions (A2)-(A4) and what we have discussed above, it yields that
$$
E\{\|U_{i1}(\ttheta,\ttheta_0)\|^2\}\leq Cdn^{-1/2}\sum_{j=1}^{n_i}\|\bW_{ij}\|^3\sum_{k=1}^K\sum_{l=1}^Kf_{ij}(\zeta_{kl})
$$
 By letting $a_i^2=Cd\sum_{j=1}^{n_i}\|\bW_{ij}\|^3\sum_{k=1}^K\sum_{l=1}^Kf_{ij}(\zeta_{kl})$ and $r=1$, Condition (B3) is obviously satisfied.

For (B4), since $A_n=C\sum_{i=1}^N\sum_{j=1}^{n_i}\|\bW_{ij}\|^3\sum_{k=1}^K\sum_{l=1}^Kf_{ij}(\zeta_{kl})=O(n)$
together with the fact that $n_i$ for $(i=1,\cdots,N)$ are bounded. Thus $A_{2n}=O(A_n)$. Condition (B4) holds.

For (B5$^{'}$),  by Assumptions (A2) and (A3), we have $E(A_n)=O(n)$. Taking the decreasing sequence of positive number $d_n$  satisfying $o(d_n)=n^{-1/2}(\log n)^4$ and $d_n=o(1)$, we can show that $\max_{1\leq i\leq N}u_i(\ttheta,\ttheta_0)=O_p(A_n^{1/2}d_n^{1/2}(\log n)^{-2})$. Since the conditions (B1)-(B4) and (B5$^{'}$) are all satisfied, then Lemma \ref{lem:bahadur} holds. $\hfill\blacksquare$ \\

\noindent
{\sc Proof of Theorem \ref{thm1}:} By Lemma \ref{lem:consistency} and \ref{lem:bahadur}, it yields that
\begin{align}\nonumber
n^{-1/2}&\sum_{k=1}^K\sum_{i=1}^N\sum_{j=1}^{n_i}\Big[\psi_{\tau_k}\{Y_{ij}-Q_Y(\tau_k;\widehat{\eeta}_{\tau_k},\widehat{t}|\bW_{ij})\}h_k(\bW_{ij};\widehat{\ttheta})-\psi_{\tau_k}\{Y_{ij}-Q_Y(\tau_k;\eeta_{\tau_k,0},t_0|\bW_{ij})\}\\\nonumber
&\times h_k(\bW_{ij};\ttheta_0)\Big]-n^{-1/2}\Big[E\sum_{k=1}^K\sum_{i=1}^N\sum_{j=1}^{n_i}\psi_{\tau_k}\{Y_{ij}-Q_Y(\tau_k;\eeta_{\tau_k},t|\bW_{ij})\}h_k(\bW_{ij};\ttheta)\Big]\Big|_{\ttheta=\widehat{\ttheta}}\\
=&o_p(1).\label{eq:thm11}
\end{align}
By applying Taylor expansion of $E\Big[\sum_{k=1}^K\sum_{i=1}^N\sum_{j=1}^{n_i}\psi_{\tau_k}\{Y_{ij}-Q_Y(\tau_k;\eeta_{\tau_k},t|\bW_{ij})\}h_k(\bW_{ij};\ttheta)\Big]\Big|_{\ttheta=\widehat{\ttheta}}$ around  $\ttheta_0$, we have
\begin{align}\nonumber
&E\Big[\sum_{k=1}^K\sum_{i=1}^N\sum_{j=1}^{n_i}\psi_{\tau_k}\{Y_{ij}-Q_Y(\tau_k;\eeta_{\tau_k},t|\bW_{ij})\}h_k(\bW_{ij};\ttheta)\Big]\Big|_{\ttheta=\widehat{\ttheta}}\\\nonumber
=&\frac{\partial\sum_{k=1}^K\sum_{i=1}^N\sum_{j=1}^{n_i}\psi_{\tau_k}\{Y_{ij}-Q_Y(\tau_k;\eeta_{\tau_k},t|\bW_{ij})\}h_k(\bW_{ij};\ttheta)}{\partial\ttheta}\Bigg|_{\ttheta=\ttheta_0}(\widehat{\ttheta}-\ttheta_0)\\\nonumber&+O_p(n(\widehat{\ttheta}-\ttheta_0)^2)\\
=&n\LLambda_n(\widehat{\ttheta}-\ttheta_0)+O_p(n(\widehat{\ttheta}-\ttheta_0)^2).\label{eq:thm12}
\end{align}
In addition, by using subgradient condition, we obtain
\begin{equation}\label{eq:thm13}
n^{-1/2}\sum_{k=1}^K\sum_{i=1}^N\sum_{j=1}^{n_i}\psi_{\tau_k}\{Y_{ij}-Q_Y(\tau_k;\widehat{\eeta}_{\tau_k},\widehat{t}|\bW_{ij}\}h_k(\bW_{ij};\widehat{\ttheta})=o_p(1).
\end{equation}
In view of (\ref{eq:thm11}), (\ref{eq:thm12}) and (\ref{eq:thm13}), we can derive the following Bahadur representation
\begin{align*}
&-n^{-1/2}\sum_{k=1}^K\sum_{i=1}^N\sum_{j=1}^{n_i}\Big[\psi_{\tau_k}\{Y_{ij}-Q_Y(\tau_k;\eeta_{\tau_k,0},t_0|\bW_{ij})\}h_k(\bW_{ij};\ttheta_0)\Big]\\
=&n^{1/2}\LLambda_n(\widehat{\ttheta}-\ttheta_0)+O_p(n^{1/2}(\widehat{\ttheta}-\ttheta_0))+o_p(1).
\end{align*}
Following from Theorem 2.2 of \cite{he1996general} together with strong consistency of $\widehat{\ttheta}$, we have
$$
n^{1/2}(\widehat{\ttheta}-\ttheta_0)=-\LLambda_n^{-1}n^{-1/2}\sum_{k=1}^K\sum_{i=1}^N\sum_{j=1}^{n_i}\psi_{\tau_k}\{Y_{ij}-Q_Y(\tau_k;\eeta_{\tau_k,0},t_0|\bW_{ij}\}h_k(\bW_{ij};\ttheta_0)+o_p(1).
$$
Finally, by applying Liapunov's central limit theorem,  $\widehat{\ttheta}$  is asymptotically normal with mean zeros and variance $\LLambda^{-1}\bH\LLambda^{-1}$. Theorem \ref{thm1} is now completed.  $\hfill\blacksquare$ \\

Based on Theorem \ref{thm1}, the following corollary holds, which is important in proving Theorem \ref{thm:RS}.
\begin{cor}\label{cor:1}
Based on the Assumptions in (A1)-(A3), we have $\widehat{\eeta}_{\tau_k}(t_0)-\eeta_{\tau_k,0}=O_p(n^{-1/2})$.
\end{cor}

\noindent
\subsection{Proof of Theorem \ref{thm:slr2}:}
The proof Theorem \ref{thm:slr1} follows the similar argument of Corollary 1 in \cite{lee2011testing}. Actually, Theorem \ref{thm:slr1} is a special case of Theorem \ref{thm:slr2} when $\Delta\beta_\tau=0$. We only need to show Theorem \ref{thm:slr2}.  Let $\mathbb{P}_n=\frac{1}{n}\sum_{i,j}(\cdot)$  be the empirical measure. Also denote $m_{\mathcal{X}}(\xxi_\tau)=-\rho_\tau(Y-\widetilde{\bX}\trans\xxi_{\tau})$ and $m_{\mathcal{X}}(\eeta_\tau,t)=-\rho_\tau\{Y-\alpha_\tau-\beta_{1,\tau}(X_{ij}-t)I(X_{ij}\leq t)-(\beta_{1,\tau}+n^{-1/2}\Delta\beta_\tau)(X_{ij}-t)I(X_{ij}>t)-\bZ\trans\ggamma_\tau\}$ as the objective function under null and alternative hypothesis, respectively.

Note that the first order derivative of $m_{\mathcal{X}}(\eeta_{\tau},t)$  evaluated at  $\widetilde{\eeta}_{\tau}$ with  $\beta_{1,\tau}=\beta_{2,\tau}$ is
\begin{eqnarray*}
\frac{\partial}{\partial\eeta_{\tau}}m_{\mathcal{X}}(\eeta_{\tau},t)\Big|_{\eeta_{\tau}=\widetilde{\eeta}_\tau}=-{\bX}(t)\left\{I(Y-\widetilde{\bX}\trans\xxi_\tau)-\tau\right\}.
\end{eqnarray*}
Thus we have
\begin{eqnarray*}
&&n^{1/2}\mathbb{P}_n\frac{\partial}{\partial\eeta_{\tau}}m_{\mathcal{X}_{ij}}(\eeta_{\tau},t)\Big|_{\eeta_{\tau}=\widetilde{\eeta}_\tau}\\
&=&n^{1/2}\mathbb{P}_n\left[ -\bX_{i,j}(t)\left\{F_{ij}(\widetilde{\bX}_{ij}\trans\xxi_\tau)-\tau\right\}\right]\\
&=&n^{1/2}\mathbb{P}_n\left[-\bX_{ij}(t)[-n^{1/2}f\{\kappa_{ij}(\tau)\}\Delta\beta_\tau(X_{ij}-t)I(X_{ij}>t)]\right]\\
&\rightarrow&n^{-1}\mathbb{P}_n\left[\bX_{ij}(t)f(\widetilde{\bX}_{ij}\trans\zzeta_{0,\tau})\Delta\beta_\tau(X_{ij}-t)I(X_{ij}>t)\right]\\
&=&\bP(t),
\end{eqnarray*}
where $\kappa_{ij}(\tau)$ lies between $\widetilde{\bX}_{ij}\trans\zzeta_{0,\tau}$ and $\widetilde{\bX}_{ij}\trans\zzeta_{0,\tau}+n^{-1/2}\Delta\beta_{\tau}(X_{ij}-t)I(X_{ij}>t)$. Similarly, we can also derive that
$$
n^{-1/2}\mathbb{P}_n\frac{\partial}{\partial\xxi}m_{\mathcal{X}_{ij}}(\xxi_\tau)|_{\xxi_\tau=\xxi_{0,\tau}}\rightarrow\widetilde{\bP}_1.
$$
Thus the limiting  distribution of $SLR_n(\tau)$ under local alternative hypothesis is
$$
1/2\left[\sup_{t\in\mathcal{T}}\left\{
\mathcal{G}(t)+{\bP}(t)\right\}^\top\mathcal{V}(t)^{-1}\left\{\mathcal{G}(t)
+{\bP}(t)\right\}-(\mathcal{G}_1+\widetilde{\bP}_1)^\top\mathcal{V}_1^{-1}(
\mathcal{G}_1+\widetilde{\bP}_1)\right].
$$
The proof of Theorem \ref{thm:slr2} is now completed. $\hfill\blacksquare$ \\
%For easier presentation, we rewrite the local alternative model (\ref{eq:local}) as
%\begin{equation}\label{eq:alt}
%Q_Y(\tau;\eeta_{\tau},t|\bW_{ij})=\alpha_\tau+\beta_{1,\tau}X_{ij}+n^{-1/2}\beta_{\tau}^*(X_{ij}-t)I(X_{ij}>t)+\bZ_{ij}\trans\ggamma_\tau
%\end{equation}
%So Models (\ref{eq:alt}) and (\ref{eq:local}) are completely equvalent. Let $\mathbb{P}_n=\frac{1}{n}\sum_{i,j}(\cdot)$ and $\mathbf{P}$ denote the empirical measure  and  the true probability measure,  respectively. Also denote $m_{\mathcal{X}}(\xxi_\tau)=-\rho_\tau(Y-\widetilde{\bX}\trans\xxi_{\tau})$ and $m_{\mathcal{X}}(\eeta_\tau,t)=-\rho_\tau\{Y-\widetilde{\bX}\trans\xxi_\tau-n^{-1/2}\beta_{\tau}^*(X-t)I(X>t)\}$ as the objective function under null and alternative hypothesis, respectively.

\noindent
\subsection{Proof of Theorem \ref{thm:RS}:}
For sake of simplicity, we assume that $b_{ij}^*\{\tau_k;\widehat{\eeta}_{\tau_k}(t_0),t_0\}$'s are independent among all subjects.
Let $\bT_n^*=(T_{n,1}^*,\cdots,T_{n,K}^*)\trans$ where $T_{n,k}^*=n^{-1/2}\sum_{i=1}^N\sum_{j=1}^{n_i}b_{ij}^*(\tau_k;\eeta_{\tau_k,0},t_0)\psi_{\tau_k}(u_{ij,\tau_k})$, $b_{ij}^*(\tau_k;\eeta_{\tau_k,0},t_0)$ is obtained  by replacing $\widehat{\eeta}_{\tau_k}$ into $\eeta_{\tau_k,0}$ in $b_{ij}^*\{\tau_k;\widehat{\eeta}_{\tau_k}(t_0),t_0\}$ and $u_{ij,\tau_k}=Y_{ij}-Q_Y(\tau_k;\eeta_{\tau_k,0},t_0|\bW_{ij})$. Then $\sum_{j=1}^{n_i}b_{ij}^*(\tau_k;\eeta_{\tau_k,0},t_0)\psi_{\tau_k}(u_{ij,\tau_k})$ are independent among  $i=1,\cdots,N$ and have mean zero. Due to the independence between subjects, we have
\begin{eqnarray}\nonumber
&&\text{Cov}(T_{n,k}^*,T_{n,l}^*)\\\nonumber
&=&n^{-1}\sum_{i=1}^N\text{Cov}\left(\sum_{j=1}^{n_i}b_{ij}^*(\tau_k;\eeta_{\tau_k,0},t_0)\psi_{\tau_k}(u_{ij,\tau_k}),\sum_{j=1}^{n_i}b_{ij}^*(\tau_l;\eeta_{\tau_l,0},t_0)\psi_{\tau_l}(u_{ij,\tau_l}) \right)\\
&=&n^{-1}\sum_{i=1}^N\bbb_{i}^*(\tau_k;\eeta_{\tau_k,0},t_0)\trans\mathcal{A}^{(kl)}_i\bbb_{i}^*(\tau_l;\eeta_{\tau_l,0},t_0),
\end{eqnarray}
where $\bbb_{i}^*(\tau_k;\eeta_{\tau_k,0},t_0)=(b_{i1}^*(\tau_k;\eeta_{\tau_k,0},t_0),\cdots,b_{in_i}^*(\tau_k;\eeta_{\tau_k,0},t_0))\trans$ and $\mathcal{A}^{(kl)}_i$ is a $n_i\times n_i$ matrix with $(j,j^{'})$ element being $\psi_{\tau_k}(u_{ij,\tau_k})\psi_{\tau_k}(u_{ij^{'},\tau_l})$ for any $k,l=1,\cdots,K$. Similar to the definition of $\bT_n^*$, we define  $\PPsi_n^*$ as a $K\times K$ matrix with $(k,l)$th element $\Psi_n^{*(kl)}=n^{-1}\sum_{i=1}^N\bbb_{i}^*(\tau_k;\eeta_{\tau_k,0},t_0)\trans\mathcal{A}^{(kl)}_i\bbb_{i}^*(\tau_l;\eeta_{\tau_l,0},t_0)$.

By using Liapunov's central limit theorem, we have $\bT_n^*\stackrel{d}{\rightarrow}N(0,\PPsi_n^*)$ and therefore $\bT_n^{*\top}(\PPsi_n^*)^{-1}\bT_n^*\stackrel{d}{\rightarrow}\chi^2_K$. Note that under Assumption (A1)-(A3), it is easy to show that
\begin{align}\nonumber
&\sup_{\|\eeta_{\tau_k}-\eeta_{\tau_k,0}\|\leq d_1(\log n/n)^{1/2}}\Big|\psi_{\tau_k}\left\{u_{ij,\tau_k}+Q_Y(\tau_k;\eeta_{\tau_k},t_0|\bW_{ij})-Q_Y(\tau_k;\eeta_{\tau_k,0},t_0|\bW_{ij})\right\}\times\\\nonumber
&\psi_{\tau_k}\left\{u_{ij^{'},\tau_k}+Q_Y(\tau_k;\eeta_{\tau_k},t_0|\bW_{ij^{'}})-Q_Y(\tau_k;\eeta_{\tau_k,0},t_0|\bW_{ij^{'}})\right\}-\psi_{\tau_k}(u_{ij,\tau_k})\psi_{\tau_k}(u_{ij^{'},\tau_k})\\\label{eq:d}
=&o_p(n^{1/4}\log n),
\end{align}
where $d_1$ is some positive constant. Thus by using Corollary \ref{cor:1} and equation (\ref{eq:d}), together with the continuous mapping theorem, we can obtain that
\begin{equation}\label{eq:pt}
\PPsi_n=\PPsi_n^*+o_p(1).
\end{equation}

It remains to show that
\begin{equation}\label{eq:bt}
\bT_n=\bT_n^*+o_p(1).
\end{equation}
To obtain desired result, it is sufficient to show $T_{n,k}=T_{n,k}^*+o_p(1)$ for any $1\leq k\leq K$. Denote $R_n(\eeta_{\tau_k})=n^{-1/2}\sum_{i=1}^N\sum_{j=1}^{n_i}\psi_{\tau_k}\{Y_{ij}-Q_Y(\tau_k;\eeta_{\tau_k},t_0|\bW_{ij})\}b_{ij}^*(\tau_k;\eeta_{\tau_k},t_0)$. Following \cite{he2000parameters} and the fact that $E[\psi_{\tau_k}\{Y_{ij}-Q_Y(\tau_k;\eeta_{\tau_k,0},t_0|\bW_{ij})\}]=0$, we obtain
\begin{equation}\label{eq:rrn}
\sup_{\|\eeta_{\tau_k}-\eeta_{\tau_k,0}\|\leq d_2n^{-1/2}}\|R_n(\eeta_{\tau_k})-T_{n,k}^*-E\{R_n(\eeta_{\tau_k})\}\|=o_p(1)
\end{equation}
where $d_2$ is some positive constant. By using Taylor expansion, we have
\begin{align}\nonumber
&E\{R_n(\eeta_{\tau_k})\}\\\nonumber
=&n^{-1/2}\sum_{i,j}E\left(b_{ij}^*(\tau_k;\eeta_{\tau_k},t_0)[\tau_k-F_{ij}\{Q_Y(\tau_k;\eeta_{\tau_k},t_0|\bW_{ij})\}]\right)\\\nonumber
=&n^{-1/2}\sum_{i,j}E\big(b_{ij}^*(\tau_k;\eeta_{\tau_k},t_0)[-f_{ij}\{Q_Y(\tau_k;\eeta_{\tau_k,0},t_0|\bW_{ij})\}\bX_{ij}(t_0)\trans(\eeta_{\tau_k}-\eeta_{\tau_k,0})\\\nonumber
&-f^{'}_{ij}\{Q_Y(\tau_k;\eeta_{\tau_k,0},t_0|\bW_{ij})\}\{\bX_{ij}(t_0)\trans(\eeta_{\tau_k}-\eeta_{\tau_k,0})\}^2+o_p(\|\eeta_{\tau_k}-\eeta_{\tau_k,0}\|^2)]\big)\\\nonumber
=&-n^{-1/2}\sum_{i,j}E\big(b_{ij}^*(\tau_k;\eeta_{\tau_k},t_0)f_{ij}^{'}\{Q_Y(\tau_k;\eeta_{\tau_k,0},t_0|\bW_{ij})\}[\bX_{ij}(t_0)\trans(\eeta_{\tau_k}-\eeta_{\tau_k,0})\}^2+o(1)]\big)\\\label{eq:rn}
=&o(1)
\end{align}
where the third ``=" holds due to the orthogonalization  projection $\sum_{i,j}b_{ij}^*(\tau_k;\eeta_{\tau_k},t_0)\bX_{ij}(t_0)=0$, and Assumption (A6) is used in the last step. Combing (\ref{eq:rrn}) and (\ref{eq:rn}), together with Corollary \ref{cor:1}, we obtain (\ref{eq:bt}). Finally, by using Slutsky's theorem,  Theorem \ref{thm:RS} holds immediately.
%To prove Theorem \ref{thm:RS}, it is sufficient to show that
%\begin{equation}\label{eq:bt}
%\|\bT_n-\bT_n^*\|=o_p(1),
%\end{equation}
%and
%\begin{equation}\label{eq:pt}
%\|\PPsi_n-\PPsi_n^*\|=o_p(1).
%\end{equation}
%For (\ref{eq:bt}), u

\renewcommand{\baselinestretch}{1.00}
\baselineskip=14pt

\bibliographystyle{apalike}
\bibliography{cqrcpm}

\end{document}